\documentclass[12pt,preprint]{aastex}

\shorttitle{The HST/ACS Coma Cluster Survey: I}
\shortauthors{Carter et al.}

\begin{document}


\title{The HST/ACS Coma Cluster Survey: I - Survey Objectives and Design\altaffilmark{1}}


\author{David Carter\altaffilmark{2},
Paul Goudfrooij\altaffilmark{3},
Bahram Mobasher\altaffilmark{3},
Henry C. Ferguson\altaffilmark{3},
Thomas H. Puzia\altaffilmark{4},
Alfonso L. Aguerri\altaffilmark{5},
Marc Balcells\altaffilmark{5},
Dan Batcheldor\altaffilmark{6},
Terry J. Bridges\altaffilmark{7},
Jonathan I. Davies\altaffilmark{8},
Peter Erwin\altaffilmark{9,10},
Alister W. Graham\altaffilmark{11},
Rafael Guzm\'an\altaffilmark{12},
Derek Hammer\altaffilmark{13},
Ann Hornschemeier\altaffilmark{14},
Carlos Hoyos\altaffilmark{12,29},
Michael J. Hudson\altaffilmark{15},
Avon Huxor\altaffilmark{16},
Shardha Jogee\altaffilmark{17},
Yutaka Komiyama\altaffilmark{18},
Jennifer Lotz\altaffilmark{19},
John R. Lucey\altaffilmark{20},
Ronald O. Marzke\altaffilmark{21},
David Merritt\altaffilmark{6},
Bryan W. Miller\altaffilmark{22},
Neal A. Miller\altaffilmark{13,23},
Mustapha Mouhcine\altaffilmark{2},
Sadanori Okamura\altaffilmark{24},
Reynier F. Peletier\altaffilmark{25},
Steven Phillipps\altaffilmark{16},
Bianca M. Poggianti\altaffilmark{26},
Ray M. Sharples\altaffilmark{19},
Russell J. Smith\altaffilmark{19},
Neil Trentham\altaffilmark{27},
R. Brent Tully\altaffilmark{28},
Edwin Valentijn\altaffilmark{25},
Gijs Verdoes Kleijn\altaffilmark{25}}

\altaffiltext{1}{Based on observations with the NASA/ESA {\it Hubble Space 
Telescope} obtained at the Space Telescope Science Institute, which is 
operated by the association of Universities for Research in Astronomy, 
Inc., under NASA contract NAS 5-26555. These obervations are associated with 
program GO10861.}
\altaffiltext{2}{Astrophysics Research Institute, Liverpool John Moores 
University, Twelve Quays House, Egerton Wharf, Birkenhead CH41 1LD, UK.}
\altaffiltext{3}{Space Telescope Science Institute, 3700 San Martin Drive, 
Baltimore, MD 21218, USA.}
\altaffiltext{4}{Dominion Astrophysical Observatory, Herzberg Institute of 
Astrophysics, National Research Council of Canada, 5071 West Saanich Road, 
Victoria, BC V9E 2E7, Canada.}
\altaffiltext{5}{Instituto de Astrof\'isica de Canarias, C/V\'ia Lactea s/n, 
38200 La Laguna, Tenerife, Spain.}
\altaffiltext{6}{Department of Physics, Rochester Institute of Technology, 
85 Lomb Memorial Drive, Rochester, NY 14623, USA.}
\altaffiltext{7}{Department of Physics, Engineering Physics and Astronomy, 
Queen's University, Kingston, Ontario K7L 3N6, Canada.}
\altaffiltext{8}{School of Physics and Astronomy, Cardiff University, The 
Parade, Cardiff CF24 3YB, UK.}
\altaffiltext{9}{Max-Planck-Institut f\"ur Extraterrestrische Physik, 
Giessenbachstrasse, D-85748 Garching, Germany.}
\altaffiltext{10}{Universit\"{a}ts-Sternwarte M\"{u}nchen, Scheinerstrasse 1,
D-81679 M\"{u}nchen, Germany}
\altaffiltext{11}{Centre for Astrophysics and Supercomputing, Swinburne 
University of Technology, Hawthorn, VIC 3122, Australia.}
\altaffiltext{12}{Department of Astronomy, University of Florida, PO Box 
112055, Gainesville, FL 32611, USA.}
\altaffiltext{13}{Department of Physics and Astronomy, Johns Hopkins 
University, 3400 North Charles Street, Baltimore, MD 21218, USA.}
\altaffiltext{14}{Laboratory for X-Ray Astrophysics, NASA Goddard Space 
Flight Center, Code 662.0, Greenbelt, MD 20771, USA.}
\altaffiltext{15}{Department of Physics and Astronomy, University of 
Waterloo, 200 University Avenue West, Waterloo, Ontario N2L 3G1, Canada.}
\altaffiltext{16}{Astrophysics Group, H.H. Wills Physics Laboratory, 
University of Bristol, Tyndall Avenue, Bristol BS8 1TL, UK.}
\altaffiltext{17}{Department of Astronomy, University of Texas at Austin, 
1 University Station C1400, Austin, TX 78712, USA.}
\altaffiltext{18}{Subaru Telescope, National Astronomical Observatory of 
Japan, 650 North A`ohoku Place, Hilo, HI 96720, USA.}
\altaffiltext{19}{Leo Goldberg Fellow, National Optical Astronomy 
Observatory, 950 North Cherry Avenue, Tucson, AZ 85719, USA.}
\altaffiltext{20}{Department of Physics, University of Durham, South Road, 
Durham DH1 3LE, UK.}
\altaffiltext{21}{Department of Physics and Astronomy, San Francisco State 
University, San Francisco, CA 94132-4163, USA.}
\altaffiltext{22}{Gemini Observatory, Casilla 603, La Serena, Chile.}
\altaffiltext{23}{Jansky Fellow of the National Radio Astronomy Observatory. 
The National Radio Astronomy Observatory is a facility of the National 
Science Foundation operated under cooperative agreement by Associated 
Universities, Inc.}
\altaffiltext{24}{Department of Astronomy, University of Tokyo, 7-3-1 Hongo, 
Bunkyo, Tokyo 113-0033, Japan.}
\altaffiltext{25}{Kapteyn Astronomical Institute, University of Groningen, 
PO Box 800, 9700 AV Groningen, The Netherlands.}
\altaffiltext{26}{INAF-Osservatorio Astronomico di Padova, Vicolo 
dell'Osservatorio 5, Padova I-35122, Italy.}
\altaffiltext{27}{Institute of Astronomy, Madingley Road, Cambridge CB3 0HA, 
UK.}
\altaffiltext{28}{Institute for Astronomy, University of Hawaii, 2680 Woodlawn 
Drive, Honolulu, HI 96822, USA.}
\altaffiltext{29}{Departamento de F\'isica Te\'orica, Universidad Aut\'{o}noma de 
Madrid, Carretera de Colmenar Viejo km 15.600, 28049 Madrid, Spain}


\begin{abstract}
We describe the HST ACS Coma cluster Treasury survey, a deep 
two-passband imaging survey of one of the nearest rich clusters of 
galaxies, the Coma cluster (Abell 1656).

The survey was designed to cover an
area of 740 arcmin$^2$ in regions of different density of both galaxies and 
intergalactic medium within the cluster. The ACS failure of January 27th 2007 
leaves the survey 28\% complete, with 21 ACS pointings (230 
arcmin$^2$) complete, and partial data for a further 4 pointings (44 arcmin$^2$).  

Predicted survey depth for 10$\sigma$ detections for optimal photometry of point sources is 
g$^{\prime}$ = 27.6 in the F475W filter, and I$_C$=26.8 mag in F814 (AB magnitudes).
Initial simulations with artificially injected point sources
show 90\% recovered at magnitude limits of g$^{\prime}$ = 27.55 and I$_C$ = 26.65. For 
extended sources, the predicted 10$\sigma$ limits for a 1 arcsecond$^{2}$ region are 
g$^{\prime}$ = 25.8 mag arcsec$^{-2}$ and I$_C$ = 25.0 mag arcsec$^{-2}$.

We highlight several motivating science goals of the survey, including
study of the faint end of the cluster galaxy luminosity function, structural
parameters of dwarf galaxies, stellar
populations and their effect on colors and color gradients,  
evolution of morphological components in a dense environment,
the nature of ultra compact dwarf galaxies, and globular cluster populations
of cluster galaxies of a range of luminosities and types. 
This survey will also provide a 
local rich cluster benchmark for various well known {\it global} scaling 
relations and explore new relations pertaining to the {\it nuclear} properties of
galaxies.

\end{abstract}



\section{Introduction and background}
\label{sec:introduction}

The Coma cluster of galaxies, Abell 1656, is along with the Perseus cluster 
the nearest rich, and dense cluster environment. Unlike the Perseus 
cluster, it is at high galactic latitude (b = 87.9$^{\circ}$) and it has been a 
popular target for study at all wavelengths. Progressively 
deeper wide-area photometric surveys of Coma have become available over the 
past 30 years, and waveband coverage has spread from the original B and V band
surveys into the near ultra-violet and infra-red (Godwin \& Peach 1977;
Godwin, Metcalfe \& Peach 1983 (GMP); Komiyama et al.\ 2002; Adami 
et al.\ 2006; Eisenhardt et al.\ 2007). A larger area around Coma is 
covered by the imaging part of Data Release 5 of the Sloan Digital Sky Survey 
(Adelman-McCarthy et al. 2007). From these surveys, samples
of galaxies have been selected for spectroscopic study, which has resulted
in an understanding of the internal dynamics of the cluster 
(Colless \& Dunn 1996;  Mobasher et al.\ 2001; Edwards et al.\ 2002; 
Gutierrez et al.\ 2004), the internal dynamics 
of cluster members (Lucey et al. 1991; Jorgensen et al.\ 1996; Smith et al.\ 2004; 
Matkovic \& Guzm\'an 2005; 
Cody et al. 2007), and their mean luminosity weighted stellar ages, 
abundances and $\alpha$-enhancement (Bower et al. 1992a,b; Guzm\'an et al.
1992; Caldwell et al. 1993; Jorgensen 1999; Poggianti et al. 2001; Moore et al. 2002; 
Nelan et al. 2005; Smith et al. 2006; Matkovic et al. 2007). Coma presents 
us with the
best opportunity to study large samples of galaxies of different luminosity,
environment and morphological type, but at a common distance, and with a common
Galactic extinction. 

There is good agreement on the distance to the Coma cluster, with independent 
studies using six different methods yielding values in the range 84 -- 108 Mpc,
as summarised in Table \ref{tab:Distances}. These values fit well with the 
current concordance cosmology: assuming $H_0$ = 71 km/s/Mpc; 
$\Omega_{\lambda}$ = 0.73; $\Omega_m$ = 0.27, and a redshift z = 0.0231 
gives a distance of 99.3 Mpc. In this paper we assume a distance 
of 100 Mpc, equivalent to a distance modulus of 35.00. 
\clearpage
\begin{deluxetable}{lccl}
\tablecolumns{4}
\tablewidth{0pc}
\tablecaption{\label{tab:Distances}The Distance to Coma determined by 
different techniques}
\tablehead{
\colhead{Technique}&\colhead{Distance (Mpc)}&\colhead{Distance Modulus}&\colhead{Reference}}
\startdata
I-band Tully-Fisher&86.3$\pm$6&34.68$\pm$0.15&Tully \& Pierce (2000)\\
K$^{\prime}$-band SBF&85$\pm$10&34.64$\pm$0.27&Jensen et al. (1999)\\
I-band SBF&102$\pm$14&35.04$\pm$0.32&Thomsen et al. (1997)\\
$D_n-\sigma$&96$\pm$6&34.90$\pm$0.14&Gregg (1997)\\
Fundamental Plane&108$\pm$12&35.16$\pm$0.25&Hjorth \& Tanvir (1997)\\
Globular Cluster LF&102$\pm$6&35.05$\pm$0.12&Kavelaars et al. (2000)\\
\enddata
\end{deluxetable}
\clearpage

\begin{figure*}
\begin{center}
\includegraphics[width=12cm]{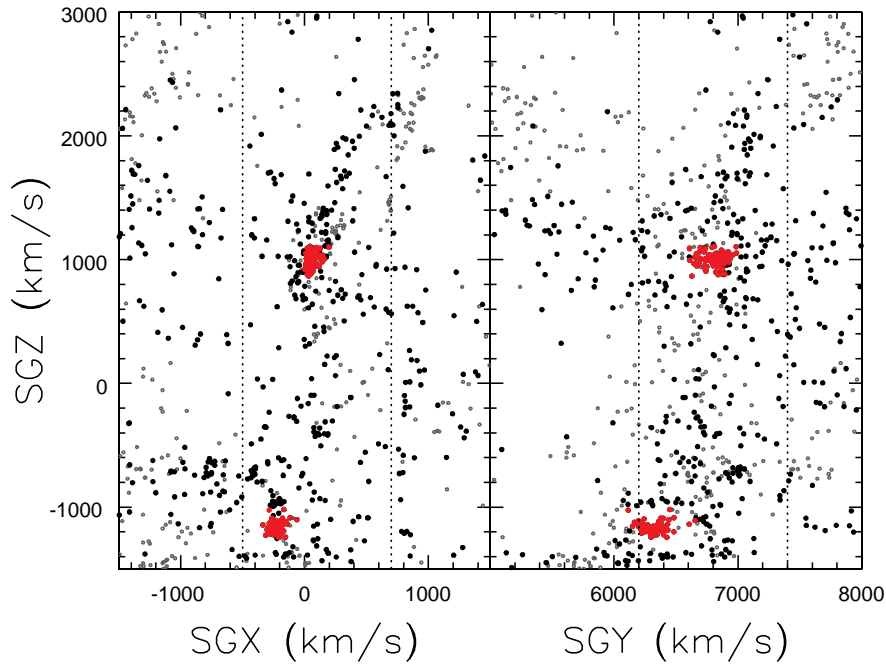}
\caption{\label{fig:coma_xyz}The location and environment of the Coma cluster
in supergalactic co-ordinates. The left panel shows a projection onto
supergalactic Y, close to the plane of the sky. The right panel
shows a projection onto supergalactic X, in this panel the 
horizontal (Y) axis is close to measured redshift. All axes are in equivalent
cz. Positions of the points are derived from measured sky 
positions and redshifts. Members of Coma (centre) and Abell 1367 (bottom) are 
plotted as filled red circles, for these galaxies the redshift used to compute 
the distance is the cluster redshift, with a random velocity offset chosen to
make the cluster appear round. For non cluster members, measured cz is used, 
and the positions are plotted as filled black circles if they have velocities
within $\pm$600 km/s of the Coma mean in the depth dimension in that 
particular panel, and as open grey circles otherwise. The Great Wall is the structure at 
SGY $\sim 6800$ km/s in the right panel and is seen face on in the left panel}
\end{center}
\end{figure*}

Coma lies in a rich region of the large-scale distribution of galaxies, at
the intersection of a number of filaments. Figure \ref{fig:coma_xyz} shows 
two projections of the distribution of galaxies in supergalactic co-ordinates 
in the region of Coma and the nearby richness class 2 cluster Abell 1367. 
The Great Wall (Geller \& Huchra 1989) a vertical structure in the two panels 
of Figure \ref{fig:coma_xyz}, runs through these two clusters,
other filaments intersect the Great Wall at the Coma cluster. 
\clearpage
\clearpage
There is a uniquely rich multi-wavelength dataset on the  Coma cluster.
X-ray observations covering a large area of the cluster have been made with 
ROSAT (White et al. 1993) and XMM-Newton (Briel et al.\ 2001), 
showing the distribution and properties of the hot intra-cluster medium (ICM), 
and X-ray properties of individual galaxies have been studied by Finoguenov
et al.\ (2004) and by Hornschemeier et al. (2006). Coma has been
shown by INTEGRAL to be an extended hard X-ray/soft $\gamma$-ray source
(Renaud et al.\ 2006). At soft X-ray
and Extreme Ultraviolet wavelengths there is a thermal excess  (Lieu 
et al. 1996; Kaastra et al. 2003; Bonamente et al.
2003; Bowyer et al.\ 2004). GALEX has been used to observe the cluster in 
the mid and near Ultraviolet  and has sufficient spatial
resolution to measure the UV properties of individual galaxies. In the 
infra-red, studies of the galaxies have been made using SPITZER,
both with IRAC at 3.6 - 8 $\mu$m (Jenkins et al.\ 2007) and with MIPS at 24 and 
70 $\mu$m (Bai et al.\ 2006). At 
radio wavelengths, VLA continuum maps cover much of the cluster (Miller
et al.\ in preparation), and in the HI 21 cm
line there are extensive VLA imaging surveys (Bravo-Alfaro et al.\ 2000, 2001),
and single-dish spectra and fluxes for samples of spiral galaxies
(Gavazzi et al.\ 2006; Vogt et al.\ 2004). 

This wealth of existing data makes Coma a prime target for studies of the 
origin and evolution of the galaxy content of clusters, and of its interaction
with the other components (gas and dark matter). Moreover, as the richest and
best studied local cluster, Coma is the zero-redshift baseline for many 
studies of high-redshift clusters (e.g. Jorgensen et al.\ 2006). 
Comparison between low- and high-redshift clusters is vital for our 
understanding of their evolution, which in turn is essential if we are to
disentangle evolutionary effects from the properties which tell us about their
formation. 

We describe here the HST/ACS Coma Cluster Treasury survey, which aims to 
provide an unparalleled database of high spatial resolution images of a sample
of cluster galaxies. At the distance of the Coma cluster
($\sim$100 Mpc), the resolution of HST/ACS (0{\arcsec}.1) corresponds to 
$\sim$50 pc. This gives essentially the same physical resolution as
ground-based observations have in Virgo and Fornax. Thus the HST/ACS Coma database
provides a valuable comparison between high- and low-density clusters, for 
studies of the effect of environment on galaxy components. 

Whilst the HST observations are the prime data upon which this survey
is based, it has already generated numerous ancillary observations with
facilities such as Subaru, Keck, MMT, UKIRT. It is concurrent with
surveys of the cluster in other wavebands, including the ultra-violet (GALEX),
infra-red (SPITZER), X-ray (Chandra and XMM-Newton) and radio (VLA).

\section{Scientific Motivation}

\subsection{The Luminosity Function}

The logarithmic slope of the low-mass end of the Cold Dark Matter ($\Lambda$CDM) mass 
function has a
slope $\alpha\approx-1.8$ (e.g. Somerville \& Primack 1999). In contrast, the 
faint end of the field-galaxy
luminosity function has a slope $\alpha\approx-1.3$, when measured either
from optically-selected surveys (Blanton et al.\ 2005), or H\,{\sc i}
(Zwaan et al.\ 2005) or $\alpha\approx-1.0$, when measured in the K-band 
(Gardner et al.\
1997; Kochanek et al.\ 2001). Luminosity functions in clusters and groups
are often not well fit by a single Schechter function (Smith et al.\ 1997), and are better
modeled by a combination of a Gaussian and a Schechter function (Ferguson
\& Sandage 1991). The composite behavior of the LF and the trends with cluster 
richness are beginning to be understood in the context of the conditional 
luminosity function (Cooray \& Cen 2005), which provides a powerful
conceptual framework for exploring the physics of galaxy formation via
studies of halos (clusters and groups) of different mass. The suppression
of the faint-end slope of the LF relative to the CDM mass function is
widely believed to be due to photoionization by the meta-galactic UV
background, which suppresses star-formation in low-mass halos. This predicts an 
environmental dependence (Tully et al.\ 2002)
because the fraction of dwarf-mass subhalos that collapse before
re-ionization is larger in higher mass halos, i.e., much larger in a 
$10^{14} M_\odot$ halo than in a $10^{11} M_\odot$
halo. This prediction can be tested directly with our survey, as the very
faint-end of the LF slope in Coma should be closer to the CDM prediction,
and at least some of the dwarfs should have very old stellar populations.

From earlier HST/WFPC2 data of a small area around NGC 4874 in the core of
the Coma cluster, Milne et al.\ (2007) determine a steep faint-end slope
($\alpha\approx-2.0$), which agrees with the slope for the faintest objects
in a small-area imaging study  by Bernstein et al.\ (1995), but not with studies
using spectroscopic redshifts (Mobasher et al.\ 2003).

Measurements of the faint-end LF slope in the Virgo cluster have been 
made by a number of authors, e.g. Trentham \& Hodgkin (2002), and 
Sabatini et al. (2007). Trentham \& Hodgkin (2002) find that the LF rises rapidly
($\alpha\approx-1.6$) between $M_B = -16$ and $M_B = -14$, but flattens faintward 
of this. The results of Sabatini et al. (2007) again suggest a steep slope
between $M_B = -16$ and $M_B = -14$, flatter between $M_B = -14$ and $M_B = -11$.
Trentham \& Tully 2002 analyse the slope of the faint end of the LF in a number
of lower density groups. They find a mean slope of $\alpha\approx-1.2$ over a
large range from $M_B = -18$ to $M_B = -10$.
Mieske
et al.\ (2007) use SBF measurements to determine cluster membership in
Fornax, and find an even flatter faint-end slope, similar to that in the local 
group ($\alpha\approx-1.1$). These authors point out however that the 
absolute number of dwarfs per unit virial mass is higher in the lower density
environments. Sabatini et al. (2007) concur that the luminosity function
slope in the field is flatter than in the Virgo cluster, although they 
parameterise this in terms of a Dwarf to Giant Ratio (DGR). These studies 
present a consistent picture of a faint-end LF slope which is strongly dependent
upon the density of the environment, and our study of the Coma LF, at a variety 
of galaxy densities, will present an important extension to this work.

The area of the survey reduces the vulnerability of determinations
of the faint-end slope to poor statistics and cosmic variance (Driver 
et al.\ 1998), and provides a sufficient range of
galaxy density to test variations in the faint-end LF across different cluster 
environments (e.g. Phillipps et al. 1998). The depth and spatial resolution
allow the use of surface brightness and 
morphological criteria to assess the probability of cluster membership 
(Trentham et al.\ 2001; Trentham \& Tully 2002) and  
examination of the relative contribution of dE, dS0 and dIrr galaxies to the faint end
LF, and of the possible evolutionary relationship between these types
(Aguerri et al.\ 2005; Mastropietro et al.\ 2005).

Different formation mechanisms give rise to the disk, bulge and bar 
components of galaxies. To understand the relative dominance of these 
physical processes
in the Universe therefore requires one to construct not simply
a galaxy luminosity function, but separate bulge, bar and disk
luminosity functions (Driver et al.\ 2007a; Laurikainen et al. \ 2005).  
The dust-corrected Millenium Galaxy Catalogue (Driver et al.\ 2007b)
provides a field galaxy comparison for the Coma component luminosity functions.

\subsection{Structure and scaling laws of dwarf galaxies}

Early-type galaxies exhibit well-defined empirical correlations between 
global galaxy parameters, such as 
luminosity, radius, surface brightness, color, velocity dispersion,
and line strength indices (Kormendy 1985; Faber \& Jackson 1976; Bender
et al.\ 1992; Guzm\'an et al. 1993; Graham \& Guzm\'an 2003). 
These scaling laws can
tell us a great deal about the physical processes operating during galaxy
formation, of which major contributors are star-formation feedback 
(Efstathiou 2000), tidal interactions (Duc et al.\ 2004), and interactions 
with the hot intergalactic medium (Babul \& Rees 1992; Roberts et al.\
2004; Tully et al.\ 2002).

Despite being the most numerous galaxy type in nearby clusters,
dE/dS0 galaxies are among the most poorly studied due to their
low surface brightness (21$<\mu_e<$25 V mag/arcsec$^2$). The scaling
laws for dwarf ($M_v > -19$) ellipticals and spheroidals are somewhat 
controversial, it is unclear whether dwarfs form a family of galaxies
distinct from giants (Wirth \& Gallagher 1984; Kormendy 1985; Caon
et al.\ 1993), or a continuous sequence with them (Caldwell 1983; 
Graham \& Guzm\'an 2003).  The ACS survey will allow
sophisticated parameterisation of the surface brightness profiles
of galaxies to  $V \sim 21$ ($M_V \sim -14$); and measurement of
basic structural parameters such as half-light radius and S\'ersic (1968)
index to $V \sim 23$ ($M_V \sim -12$). Such limits are similar to those
for measuring velocity dispersions and line strengths with large ground-based
telescopes. Substantial samples from multi-object spectrographs on 4 metre 
class telescopes down to $B \sim 19$ already exist (see 
section \ref{sec:introduction}).

The study of the scaling laws in various environments also provides the
key observational reference needed to test specific predictions of  
current theoretical models of dwarf galaxy formation and evolution. For  
instance, the galaxy harassment model predicts that dE/dS0s in the 
highest-density cluster regions should have steeper light profiles, the 
fraction of nucleated dE/dS0s should be higher, the fraction of any remaining 
disk structure should be lower, and they should have higher metallicities  
than those located in lower-density cluster environments (Moore et al.\
1998).

\subsection{The effect of environment upon morphological components}

The high density in the Coma cluster core makes it an ideal place to
investigate the morphology-density relation, in which the average
bulge-to-disk ($B/D$) luminosity ratio increases with galaxy density 
(Dressler 1980). This is understood to result from transformation 
of spirals into early types through processes such as 
harassment (Moore et al.\ 1996). An example of this transformation in progress
may be the Coma cluster galaxy GMP 3629 (Graham, Jerjen, \& Guzm\'an 2003).
Furthermore, there is a strong radial gradient in HI
content in the spirals, those in the cluster core being severely depleted
(Bothun et al.\ 1984; Bravo-Alfaro et al.  2000).  This, together with similar
results in Virgo, supports the notion of gas stripping and the subsequent
cessation of star formation (Conselice et al.\ 2003).  From a quantitative
morphological analysis using high-quality ground-based images, Gutierrez et
al.\ (2004) find that disks in the Coma core are 30\% smaller than in the
field for a given bulge size.

Bulge-disk decomposition in the presence of nuclear components
depends critically on spatial resolution (Balcells et al.\ 2003).  The high
resolution of ACS allows identification of disks, unobscured primary and nuclear
bars, classical bulges, resonance starburst rings, compact disks, disky
bulges, and pseudo-bulges (e.g., Kormendy \& Kennicutt 2004; Athanassoula
2005).  While observations in the rest-frame $I$-band (F814W) may miss some
highly obscured morphological features, a comparison of the optical properties
across field and dense clusters will set important constraints on how
environment influences the morphology and structure of galaxies.  
The radial dependence of bulge and disk morphologies can constrain the 
roles of mergers (Aguerri et al.\ 2001) and of disk truncation processes.

\subsubsection{Stellar Bars and Disks}

Stellar bars redistribute angular momentum in disk galaxies, driving their
dynamical and secular evolution (Erwin et al.\ 2003; Kormendy \& Kennicutt
2004; Jogee et al.\ 2004, 2005; Erwin 2005).  Many barred galaxies host
molecular gas concentrations of up to several thousand $M_{\tiny \sun}$
pc$^{-2}$ in the inner kpc, intense circumnuclear starburst rings, and disky
bulges (e.g., Jogee 1999; Jogee et al.\ 2005).  Bars are abundant in the field 
out to $z \sim$~1.0, corresponding to the last 8 Gyr (Elmegreen et al.\ 2004;
Jogee et al.\ 2004; Zheng et al.\ 2005), however we know much less about barred
disks in dense clusters.  Varela et al.\ (2004) note that
bars are twice as frequent in perturbed galaxies as in isolated galaxies, and
Elmegreen et al.\ (1990) note a higher frequency of bars in binary galaxy
systems compared with isolated galaxies, so there is some evidence for
variation with environment.  The closest to a systematic environmental survey
is that of van den Bergh (2002), who found the frequency of bars in
spiral galaxies in clusters and in the field to be similar, although he did not 
probe the richest cluster environments.

Bars and similar structures  (e.g., nuclear bars, rings, and
spirals) are detectable in HST images if they have
diameters $\sim 3$--5 FWHM (Sheth et al.\ 2003; Jogee et al.\ 2004; Lisker et
al.\ 2006).  At the resolution of the Coma survey they will be
detected down to sizes of $r \sim 150$ pc.  This will enable 
identification and characterization of unobscured large-scale bars across the 
Hubble sequence (e.g., Erwin 2005), and across different environments.
In moderately inclined 
galaxies, the survey can identify compact disks and disky ``pseudobulges'' 
(Erwin et al.\ 2003; Kormendy \& Kennicutt 2004; Athanassoula 2005; 
Jogee et al.\ 2005).  Coma is an excellent case study for the 
significance of inner/nuclear components in S0 and spiral galaxies, and a key 
reference for comparison with studies of nuclear bar and ring frequencies in 
local field samples (e.g. Erwin \& Sparke 2002; Laine et al.\ 2002; 
Knapen 2005).

The surface brightness profiles of
stellar disks fall into three classes (Erwin, Beckman, \& Pohlen 2005; Pohlen
\& Trujillo 2006; Erwin, Pohlen, \& Beckman 2007): classic single-exponential
profiles, downward-bending ``type II'' profiles (Freeman 1970; ``truncations''), 
and upward-bending ``type III'' profiles
(``antitruncations'').  Type II profiles are common in early-type barred
galaxies, and appear linked to the Outer Lindblad Resonance of bars; type III
profiles, on the other hand, appear anti-correlated with bars, and may be a
signature of minor mergers (Younger et al.\ 2007).  A comparison of barred
S0--Sb galaxies in the Virgo Cluster and the local field shows a dramatic
dichotomy: type II profiles are common in the field but essentially absent in
Virgo (Erwin 2007, in prep).  If this can be confirmed for the Coma Cluster,
it points to a clear role for the cluster environment in shaping the outer
disks of S0's and early-type spirals, and suggests an important test for
distinguishing models of S0 formation in clusters from the field.

\subsection{Colour gradients and stellar populations}

The steepness of the radial color gradients of dwarf and giant galaxies directly
reflects their merging history: monolithic collapse imposes an initially
steep negative metallicity gradient (Carlberg 1984), which will be
progressively diluted by subsequent major mergers (Bekki \& Shioya 1999;
Kobayashi 2004). Semi-analytic models of hierarchical galaxy
formation predict differences in the merger history as a function of
galaxy mass and environment (Cole et al.\ 2000). The Coma survey
provides an unbiased sample of cluster dwarfs, so that the global 
scaling relations of different dwarf subtypes (dE, dS0, dE-N, dIrr) can be 
reliably determined. Coma ellipticals have metallicities that seem to 
correlate with galaxy mass, which is broadly consistent with monolithic 
collapse models 
(e.g. Forbes et al.\ 2005). In contrast, dwarf ellipticals and spheroidals
display a variety of color gradients, even positive in some places, 
which is an indication of recent star formation (van Zee et al.\ 2004). 

Complemented by $K$-band and IFU observations, the ACS images can be 
used to interpret the observed color distributions in terms of ages and 
metallicities, using photometric techniques (James et al. 2006) and line indices
(e.g. Poggianti et al.\ 2001; Mehlert et al.\ 2003; S\'anchez-Bl\'azquez et al.\
2006).

\subsection{Globular clusters and UCDs}

The Coma cluster hosts several bright early-type galaxies, which are known
to have very rich globular cluster (GC) systems (Ashman \& Zepf 1998), and
many less massive galaxies with their own star cluster systems that
were assembled and evolved in the dense Coma cluster environment. 
There have been HST/WFPC2 studies of the
GC systems of a small number of bright Coma ellipticals: IC~4051 (Baum et al.
1997; Woodworth \& Harris 2000), NGC~4881 (Baum et al. 1995) and the cD
galaxy NGC~4874 (Kavelaars et al.\ 2000; Harris et al.\ 2000).
Mar\'in-Franch \& Aparicio (2002) find a wide range of specific frequency
$S_N$ values among the brighter galaxies in Coma, using the Surface
Brightness Fluctuation (SBF) technique on ground-based data.

The ACS Virgo cluster survey has provided a number of important
new results on the GC populations of early-type galaxies in the sparser
Virgo cluster  (e.g.~Jord\'an et al.\ 2005; Peng et al.\ 2006a;
Mieske et al.\ 2006). At the distance of Coma it is only possible to
resolve half-light radii 
down to $\sim\!20$ pc (0\arcsec.04). Moreover, the peak of the Globular Cluster
Luminosity Function (GCLF) in Coma is suspected to be at $V = 27.88 \pm
0.12$ (Kavelaars et al.\ 2000), some 1.3 magnitudes fainter than our 10-$\sigma$
photometric completeness limit for point sources. Nevertheless, if the
GCLF has a Gaussian form with $\sigma$ = 1.4 magnitudes some 18\% of the
GCs are expected to be brighter than this limit, allowing us to study the
color distributions, color-magnitude diagrams and spatial distribution of
GCs in a much wider range of galaxies than in the targeted Virgo cluster
survey (from $M_V$ = -23.3 to -15), to a larger physical radius from the galaxy 
centers (up to 100 Kpc). Stacking the GC systems of galaxies in bins will allow 
study of their properties as a function of host galaxy type, luminosity and 
environment, and thus of the relationship between GC systems and their hosts,
and the processes which lead to the formation of the rich GC systems of the massive
galaxies (Pipino et. al.\ 2007)

\subsubsection{The spatial distribution of clusters and the epoch of reionization}

The old, metal-poor GC population in giant ellipticals may 
provide the best available tracer of their dark halos, and thus may be useful 
in testing the predictions of $\Lambda$CDM models. Moore et al.\ (2006) and 
Bekki \& Yahagi (2006) find that the final radial distibution of the old GCs 
depends upon the redshift of 
truncation of GC formation, which in turn might depend upon the epoch of
reinonization. If the truncation is earlier, then the final radial GC 
distributions will be steeper and more compact (Bekki \& Yahagi 2006).

Bassino et al.\ 
(2006) present a study of the globular clusters around NGC 1399, the dominant
galaxy in the Fornax cluster. They note a good agreement with a projected
dark matter NFW (Navarro et al.\ 1996) density profile. Abadi et al.\ (2006) 
test the radial  density profile of GCs around the somewhat isolated galaxies 
M31 and the Milky
Way, comparing with the numerical simulations. They report that such luminous
halos are similar in shape to their dark matter counterparts. On the other hand
Merritt et al.\ (2006) in a careful analysis of a set of $\Lambda$CDM 
simulations, find that the halo density distributions are better fit by
the much earlier model of Einasto (1965, 1972, 1974) than by NFW or a number of
other alternative models.

\subsubsection{UCDs and transition objects}

Ultra-compact dwarfs (UCDs), at $-13.5 \leq M_V \leq -11.5$, have been 
detected from the ground in the Fornax
and Virgo Clusters (Hilker et al.\ 1999; Drinkwater et al.\ 2000; 
Phillipps et al.\ 2001; Jones et al.\ 2006). HST imaging 
of the bright Fornax UCDs shows that they have half-light radii of $10\!-\!20$
pc, larger than both Local Group GCs and typical dwarf galaxy nuclei, but smaller than 
any previously known dwarf galaxies (De Propris et al.\ 2005). Similar but fainter 
objects have since been found (Ha\c{s}egan et al.\ 2005 - their dwarf galaxy globular 
cluster transition objects, DGTOs). These overlap in luminosity with, the 
brightest GCs, leading to the question of whether UCDs and large
GCs are indeed fundamentally different objects. UCDs and DGTOs appear to be a feature of
denser environments, so it is important to quantify the population of this
new type of object in Coma and compare it with the less rich Virgo and Fornax.

Proposed scenarios see UCDs (see Hilker 2006, for a review) as the tidally
`threshed' remnant nuclei of former nucleated dwarf elliptical
galaxies in the cluster core (Bekki et al.\ 2001), the similar remains of
late-type spirals with nuclear star clusters (Moore et al.\ 1998), more
massive, perhaps intra-cluster versions of ordinary GCs (Hilker et
al.\ 2004), merged (super)star clusters (Fellhauer \& Kroupa 2002),
products of massive starbursts during the merger of two gas-rich galaxies
(Fellhauer \& Kroupa 2005), or even left-over primordial building blocks
of central galaxies in the dense galaxy cluster environment (Rakos \& Schombert 
2004).

The point source magnitude limit of the Coma survey is deep enough to
observe compact objects down to $M_V \simeq -9$ mag, so we can detect both
UCDs and DGTOs down to GC
luminosities. At the distance of Coma, the half-light radii of the brighter examples
will be 0.02 to 0.04 arcsec,
corresponding to $\sim\!10-20$ pc (Drinkwater et al.\ 2003). These sizes are 
similar to those of extended star clusters in the Virgo Cluster which have been 
successfully differentiated from point sources by Peng et al.\ (2006b) using ACS. 
They can be reliably distinguished from background
galaxies at magnitudes $V \sim 22$ to 25 mag which have half-light radii
typically between 0.2 and 1.0 arcsec (e.g. Roche et al.\ 1997).

\subsection{The nature of ``E+A'' galaxies}

Spectroscopic surveys (Caldwell et al.\ 1993;
Poggianti et al.\ 2004) have identified several 
galaxies with post-starburst spectra in both the core and the infall region of 
Coma. Suggestions as to what triggered these bursts range from equal-mass 
galaxy mergers (Barnes \& Hernquist 1992) to high-speed impact with shock 
fronts 
of infalling substructures (Poggianti et al.\ 2004) or generally with the dense 
ICM (Tran et al.\ 2005). The key to distinguishing
between these processes is the spatial distribution of the intermediate-age 
populations. It would be concentrated to the center if major mergers were 
involved (Barnes \& Hernquist 1992) while it would be either an extended or 
off-centered phenomenon if the galaxy had experienced an interaction with 
the ICM.  The Coma survey includes a number
of ``E+A'' or ``k+a'' galaxies at a range of luminosity, and the 
morphological features and color maps
generated from the two ACS passbands will provide a valuable diagnostic
of the physical origin of the ``E+A'' phenomenon.

\subsection{Nuclear Star Clusters and Central Massive Objects}

Massive black holes at the centers of spheroidal stellar
systems correlate with the large scale properties of the host
spheroid.  Popular correlations involve: (i) spheroid velocity dispersion
(Ferrarese \& Merritt 2000; Gebhardt et al.\ 2000); (ii) stellar concentration
(Graham et al.\ 2001; Graham \& Driver 2007); (iii) luminosity (McLure \&
Dunlop 2002; Marconi \& Hunt 2003; Graham 2007); and (iv) mass (Marconi \&
Hunt 2003; H\"aring \& Rix 2004).

The relationship between nuclear star clusters and central massive objects
is of increasing interest, although the formation of the former remains poorly understood. 
They are observed in about 80\% of
intermediate-luminosity, early-type galaxies (e.g., Ferguson \& Binggeli 1994;
Graham \& Guzm\'an 2003; Cot\'e et al.\ 2006; Jordan et al.\ 2007; Balcells et
al.\ 2007). Graham \& Guzm\'an (2003) find that the luminosity 
of nuclear star clusters in dEs correlates strongly with the luminosity of the host
spheroid.  This luminosity trend has also been shown to exist in the bulges of
lenticular and early-type spiral galaxies (Balcells et al.\ 2003).
It has also been proposed that they follow the same 
$M_{\rm nucleus}$--$M_{\rm spheroid}$ relation as defined by supermassive black holes 
(Ferrarese et al.\ 2006b; Balcells, Graham \& Peletier 2007), and the same 
$M_{\rm nucleus}$--Sersic index relation (Graham \& Driver 2007).

Graham \& Driver (2007) suggest that the driving physical relation may 
be with the central stellar density (prior to core depletion) rather than 
a {\it global} property of the host spheroid. Given the known trend
between central stellar density and host spheroid luminosity (e.g., Graham \&
Guzm\'an 2003; Merritt 2006a), the relations with global properties
may be secondary in nature. It is therefore important to examine
the connection between the mass of the nuclear star
clusters and the central stellar density (projected, $\mu_o$, and deprojected,
$\rho_0$) of the host spheroid. 

Study of the luminosity function, color-magnitude relation, relationship
between nuclear and host galaxy color, and the radial distribution within
the cluster (Ferguson \& Sandage 1989; Lisker et al.\ 2007), and comparison 
with the lower density Fornax and Virgo environments, may help shed
some light on potential formation mechanisms. Measurement of spatial offsets 
between nuclear clusters and the outer isophotes may reveal an oscillation of the 
nucleus about the center of the potential, which will be of larger amplitude
in shallower potentials (Binggeli et al.\ 2000). Furthermore, apparant double nuclei 
may in some cases be a sign of an edge-on nuclear disk
(Kormendy 1988; Tremaine 1995), from which we may be able to determine the
presence of a central massive black hole (e.g.\ Debattista et al.\ 2006).

Giant elliptical galaxies display cores which are partially depleted of 
stars.  A possible mechanism is the wrecking ball action of supermassive black 
holes (SMBHs) from the progenitor
galaxies as they sink to the center of the newly-formed galaxy during a merger
(Begelman et al.\ 1980; Ebisuzaki et al.\ 1991, Merritt et al.\ 2007).
Cores may also be enlarged when black holes are ejected from
galaxy centers by the gravitational wave rocket effect following
coalescence of a SMBH binary (Redmount \& Rees 1989; Merritt et al.\ 2004; 
Gualandris \& Merritt 2007).

Using the core-S\'ersic model (Graham et al.\ 2003; Trujillo et al.\ 2004) it
is possible to quantify the sizes (and mass deficits, $M_{\rm def}$) of these
cores, and to predict each galaxy's central black hole mass, $M_{\rm bh}$, 
using the measured S\'ersic index n (e.g., Graham \& Driver 2007).
Given that $M_{\rm def}/M_{\rm bh}$ scales roughly linearly with $N$,
where $N$ is the number of major dry mergers, such measurements can be used to place
constraints on the dry merger history of such cluster galaxies (Graham 2004;
Ferrarese et al.\ 2006a; Merritt 2006b), and to constrain $N$ as a function of galaxy 
magnitude.

\section{Survey Design}

\subsection{Survey area}

The ACS Wide Field Camera (ACS/WFC) has a field of view of 11.3 square 
arcminutes. The camera contains two 4096 x 2048 pixel
CCDs, with an inter-chip gap of some 3 arcseconds. The capabilities of ACS at 
the time of our pre-anomaly observations are descrived in some detail by 
Pavlovsky et al.\ (2006). 

The aim of the survey is to provide a large sample of galaxies for study in a 
high density environment, and at the same time a comparison sample in a lower 
density region of the cluster for the specific science goals of examining the
effect of environment upon morphology, structural parameters and stellar
populations. Because of the density of confirmed cluster galaxies in the core 
(e.g. Colless \& Dunn 1996) we have adopted a tiling strategy in the core 
region, tiling a region of approximately 18 x 21 arcminutes, using 42 ACS
pointings in a 7 x 6 pattern, with some overlap
between adjacent tiles. An HST orientation of 282 degrees was chosen for these 
tiles in order to maximise the time for which the obeservations could be 
scheduled in two gyro mode (Sembach et al.\ 2006). A tile of the central 
mosaic was moved to the southern edge of the mosaic area, away from the 
star HD112877 (V=7.17) which would have a negative impact on ACS observations 
near its position. 

In the outer part of the cluster the density of known members is less
than one per ACS tile, and we decided to target known members which can be 
used to address some of our primary science goals. The best studied region
outside the core is the infall region around NGC4839 (Neumann et al.\ 2001)
where there have been a number of photometric surveys (Komiyama et al.\ 2002)
and spectroscopic surveys for poststarburst galaxies (Caldwell et al.\ 1993),
and line strengths and velocity dispersions of dwarf spheroidals (Poggianti
et al. 2001; Matkovic \& Guzm\'an 2005, 2008; Cody et al.\ 2007). 
40 further ACS 
pointings were defined, each containing one, or in most cases more, cluster
members from these spectroscopic surveys. In some cases the orientation was 
left free in the HST Phase 2 submission, in others a range was defined in
order to ensure that more than one target galaxy was included. Table 
\ref{tab:Visits}
lists the positions and orientations of our survey tiles. In this Table
Column 1 gives the HST Visit number within the program; column 2
the field title. Columns 3 and 4 give the field center RA and Dec, and 
column 5 the field orientation of the observation, or the orientation specified 
for visits which have not yet been completed. Column 6 lists the
spectroscopically confirmed cluster members within the field, with identifiers: 
GMP from Godwin et al.\ (1983); K from Komiyama et al.\ (2002). Column 7 gives 
the date of the observation, and column 8 the number of dither positions 
which have been observed by the time of the January 2007 ACS failure. At this time,
21 of our 82 fields have been fully observed, and further 4 have either two 
or three dither positions observed. The survey is thus 28\% complete.
 
\begin{figure*}
\begin{center}
\includegraphics[width=8cm]{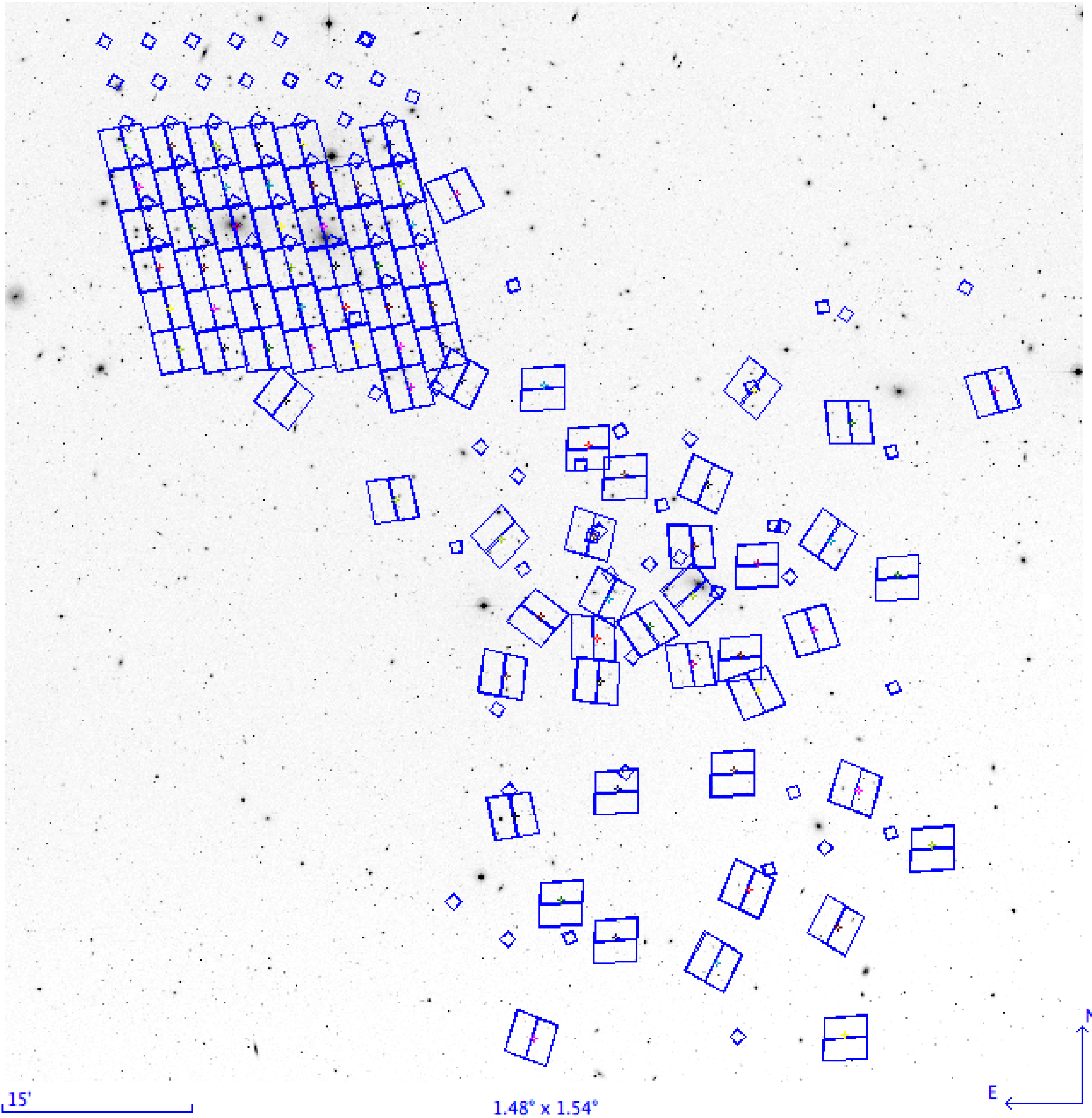}
\includegraphics[width=8cm]{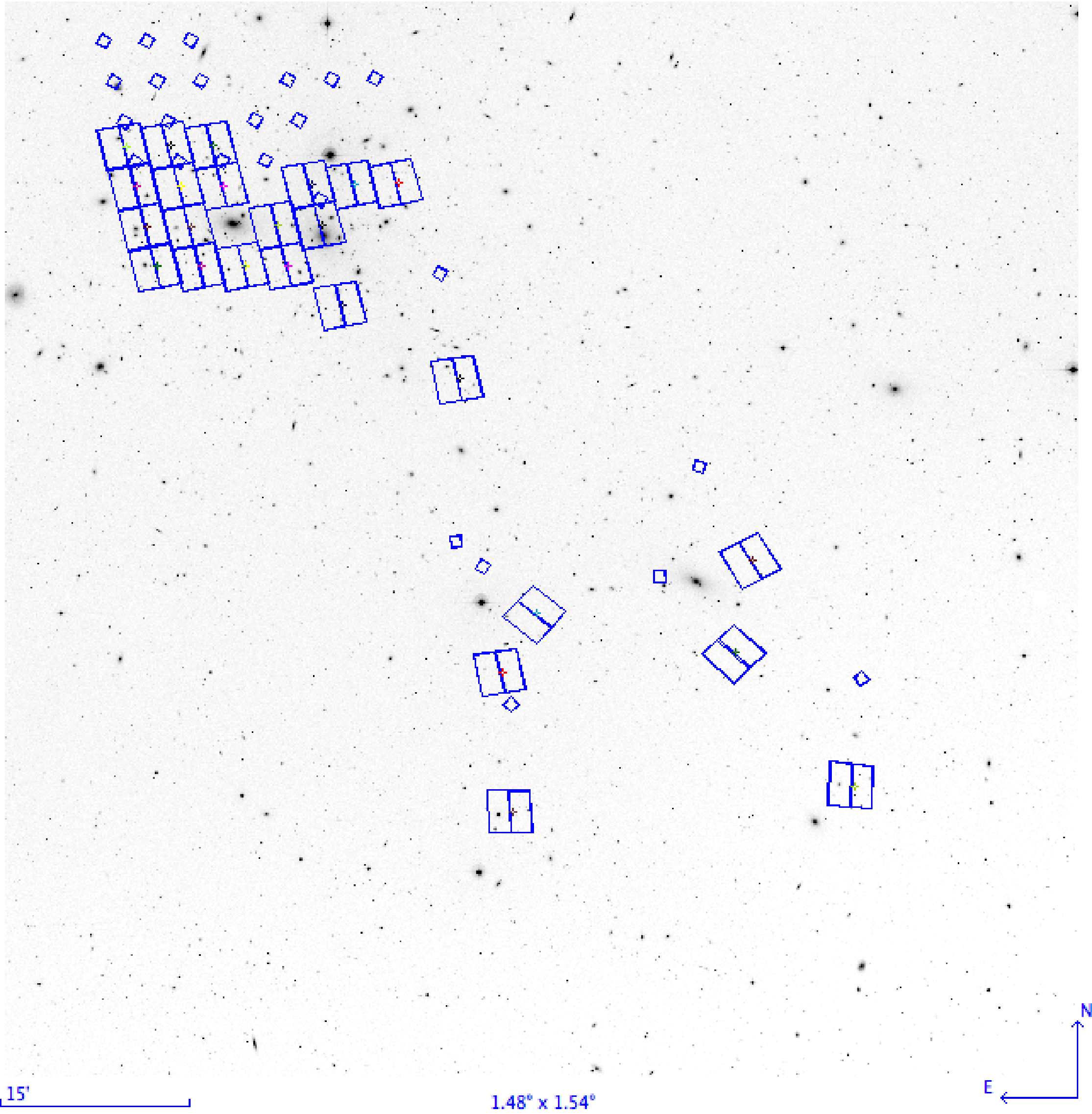}
\caption{\label{fig:surveytiles}Left: Positions of ACS tiles for the survey 
proposed, overlayed
on a DSS image of the core and infall region of the Coma cluster. The small
squares represent the fields of NICMOS parallel exposures. Right: The survey 
as at the ACS failure of January 2007. Tiles shown have some or all of the 
proposed observations complete by this date.In this panel requested 
orientations are replaced by observed orientations if they differ.}
\end{center}
\end{figure*}

Figure \ref{fig:surveytiles} shows in the left panel the 
positions and orientations of the survey tiles, and the locations of NICMOS 
parallel observations. For cases where a range of orientations are specified
in the Phase 2 submission the orientation shown is the midpoint of the range, 
and those tiles with free orientation are shown at 90 $^{\circ}$. The positions
and final orientations of the 25 fully or partially observed fields are shown 
in the right hand panel of this figure.
\clearpage
\clearpage

\subsection{Choice of passbands, exposure times and dither parameters}

\begin{figure}
\begin{center}
\includegraphics[width=5.5cm,angle=270]{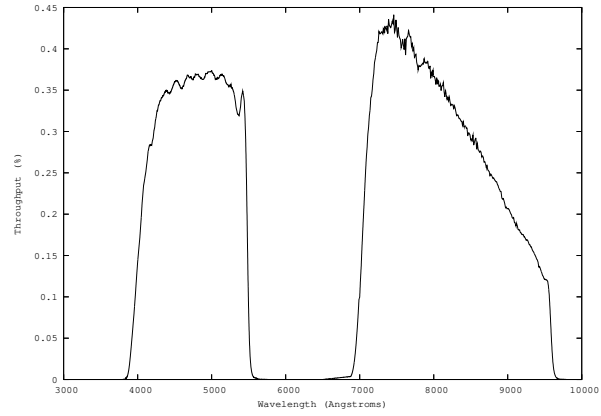}
\caption{\label{fig:throughput}Total throughput curves of ACS 
CCDs plus filter for F475W passband
(left) and F814W passband (right), after Sirianni et al. (2005)}
\end{center}
\end{figure}

Color information is essential to some of the most important aims of the survey,
so we require two passbands. For these passbands we choose F814W, as the 
passband which will give the deepest data for structural and luminosity 
function studies, and F475W, which is a compromise between color baseline and 
speed. These passbands are close to Cousins I and SDSS g' respectively. 
Transformations from these ACS/WFC passbands to standard Johnson/Cousins
passbands are given by Sirianni et al.\ (2005). Total system throughput as a 
function of wavelength for all ACS/WFC filters, from the work of Sirianni 
et al.\ (2005), is given on the ACS project website 
\footnote{http://adcam.pha.jhu.edu/instrument/photometry/} and we plot in 
Figure \ref{fig:throughput} the throughput for our two passbands.
\clearpage
\clearpage
We defined a dithering pattern with a large (3 arcsecond) dither across the 
inter-CCD gap on ACS/WFC, plus a sub-dither to remove hot pixels, resulting
in four positions over the two orbits. The large move was made between orbits.
Each passband was observed at each dither position. The large dither point 
spacing was 3.011 arcseconds at a pattern orientation of 85.28$^{\circ}$, and 
the hot pixel sub-dither spacing was 0.2412 arcseconds at a pattern orientation
of 22.3239$^{\circ}$. POS-TARG equivalent positions are given in Table
\ref{tab:POS-TARG}. 

\clearpage
\begin{deluxetable}{lll}
\tablecolumns{3}
\tablewidth{0pc}
\tablecaption{\label{tab:POS-TARG}POS-TARG equivalent of dither pattern used}
\tablehead{
\colhead{Dither Position}&\multicolumn{2}{c}{POS-TARG}}
\startdata
1&0.0&0.0\\
2&0.222&0.092\\
3&0.247&3.001\\
4&0.469&3.093\\
\enddata
\end{deluxetable}
\clearpage

Exposure times were 350 seconds in F814W and 640 seconds in F475W at each 
dither position. F814W frames are somewhat deeper than F475W, but it was not 
possible to distribute the time more towards F475W, as this would have 
resulted in F814W frames too short to dump the ACS buffer, and thus a 
substantial increase in overheads. In some visits the F475W exposure time
for the final dither position could be slightly longer, due to savings in
the reacquisition overhead (as compared with the original acquisition), 
resulting in a total F475W integration time of between 2560 and 2680 seconds.

In Table \ref{tab:maglimits} we present magnitudes and surface brightnesses
for signal-to-noise ratio (S/N) = 10, calculated with the ACS Exposure Time 
Calculator (ETC). In these calculations the ETC is set to assume standard
background conditions, and an elliptical galaxy spectral energy distribution. 
The ETC gives two values for S/N for a point source: that in the ``default''
aperture, which for ACS/WFC is a circle of 0.2 arcsec radius; and an
``optimal'' S/N, assuming that a PSF fitting algorithm is used. In 
columns 3 and 4 of Table \ref{tab:maglimits} limits are presented in each 
passband for each of these S/N values to reach 10.

The real completeness and reliability limits, and 
contamination with spurious sources near these magnitude limits will depend
upon real background conditions, to which the halos of large galaxies will 
contribute, and upon real HST guiding performance. For point sources we have 
estimated the limits by injecting artificial point sources into 
our ACS frames from one of our outer fields, and recovering them with 
SExtractor. In column 5 of Table \ref{tab:maglimits} we present the magnitude
at which 90\% of artificial injected sources are recovered. This limit is, as 
expected, between the two calculated limits, but encouragingly close to
the ``optimal S/N'' calculated limits. 

For galaxies the limits will depend upon surface brightness and structural 
parameters. These will be discussed further in Paper II in this series 
(Hammer et al.\ 2008).

In column 6 of Table \ref{tab:maglimits} we present a calculated limit
in mag arcsec$^{-2}$, obtained (using the ETC) for a S/N = 10 for a uniform 
surface brightness region of 1 arcsec$^2$. 

All magnitudes and surface brightnesses have been converted from the Vega
magnitude system used by the ETC to AB magnitudes, using the AB magnitude
of Vega in the chosen filters (e.g. Sirianni et al. 2005). 
\clearpage
\begin{deluxetable}{llllll}
\tablecolumns{4}
\tablewidth{0pc}
\tablecaption{\label{tab:maglimits}Columns 3 and 4 - predicted 10$\sigma$ limits for 
optimal extraction of point sources; column 5 - limit for 90\% recovery 
of injected point sources; column 6 predicted  
surface brightness (1 sq arcsec) limits; all in the AB magnitude system.}
\tablehead{
\colhead{Passband}&\colhead{Exposure}&\multicolumn{2}{c}{ETC Limit (S/N=10)}&\colhead{Measured}&\colhead{SB (mag arcsec$^{-2}$)}\\
\colhead{}&\colhead{(seconds)}&\colhead{``Default''}&\colhead{``Optimal''}&\colhead{90\% recovery}&\colhead{S/N=10 in 1 arcsec$^2$}}
\startdata
F475W&2560&g$^{\prime}$ = 26.75&g$^{\prime}$ = 27.6&g$^{\prime}$ = 27.55&g$^{\prime}$ = 25.8 \\
F814W&1400&I$_C$ = 25.95&I$_C$ = 26.8&I$_C$ = 26.65&I$_C$ = 25.0\\
\enddata
\end{deluxetable}
\clearpage

\section{ACS Data Reduction}

The ACS data processing was performed at STScI. It involved a dedicated
pipeline based on a wrapper script using PyRAF/STSDAS modules that performed
CCD data reduction and cosmic ray cleaning, as well as the combination
of individual dithered images using {\tt MultiDrizzle} (Koekemoer et al.\ 2003), 
which makes use of the {\tt Drizzle} software  (Fruchter \& Hook 2002)
to remove geometric distortion and map the input exposures onto a rectified
output frame. We describe details of the ACS data reduction below.     

\subsection{Initial CCD Reduction}

As the observations of each {\it HST\/} visit were obtained, the data were run
through the STScI on-the-fly-reprocessing (OTFR) pipeline for ACS data. This
pipeline reduction involves basic CCD reduction by means of the IRAF/STSDAS
task {\tt calacs}, which performs bias subtraction, gain correction, dark
subtraction, flat fielding, and the identification of bad pixels. Finally, the
OTFR pipeline identifies the exposures taken through each filter (F475W
vs.\ F814W) and combines them using the OTFR version of {\tt MultiDrizzle} for
ACS data. These first-pass pipeline products were used for quicklook
purposes. Final images were created by a second-pass calibration
procedure which contained the steps described below. 

\subsection{Updated Reference Files}

A few weeks after the each visit's data were observed, we downloaded
more accurate calibration reference files from the {\it HST\/}
archive, in particular the up-to-date superbias and superdark
reference files which were created from bias and dark exposures
obtained contemporaneously with the science data (see,
e.g., Lucas et al.\ 2006). These files were used for the final calibrations. 

\subsection{Bias Level Offsets}

In virtually all ACS datasets of our Coma survey, we encountered significant
bias level offsets between the four quadrants of the {\tt MultiDrizzle} output
images. These offsets, which ranged from a few tenths to several Data Numbers,
are due to randomly varying differences between the bias level in the parallel
overscan region of a given ACS/WFC CCD chip and the level in its active region
(see, e.g., Sirianni et al.\ 2003; note that ACS/WFC contains four CCDs that 
are read out by their own amplifiers). A script was developed to measure these
bias offsets between the four quadrants using iterative statistics on a large
number of pairs of areas located close to (and symmetrically relative to) the
quadrant-to-quadrant borders. The measured offsets were subtracted from the
appropriate quadrants of the superbias reference file being used in the final
image reduction of each set of exposures with a given filter.   

\subsection{Image Registration}

In many cases, the spatial registration of images taken during each {\it HST\/}
visit of this Coma treasury program was less accurate than expected, even
though all images were planned to be taken without any change of prime guide star,
telescope position or roll angle. This problem was due to two compounding
circumstances: {\it (i)\/} several visits were forced to use single-star
guiding due to unforeseen problems with the {\it HST\/} Guide Star Catalog
version 2 in the region of the Coma cluster, {\it (ii)\/} the observations
of our program unearthed a software bug in the ground system of the two-gyro
guiding mode of {\it HST\/} for guide star re-acquisitions when single guide 
stars were used. The result of these issues was that significant spatial
offsets are present between images (and their world coordinate system (WCS) 
header keywords) taken in different {\it HST\/} orbits, even if taken during
one and the same visit. Since it is crucial to align images to better than
$\sim 5 - 8$~milliarcseconds (i.e., $\la 0.15$ ACS/WFC pixel) in order to
achieve accurate cosmic ray rejection among the separate exposures within a
visit and to retain satisfactory point spread functions (PSFs) in the
combined image, we registered all images within a visit to a common
astrometric grid as follows.   

First, we ran {\tt MultiDrizzle} on every set of associated images (i.e.,
images taken in one visit with the same filter) using the {\tt driz\_separate =
  yes} setting. This created so-called ``singly drizzled'' images, which are
the individual input science images in the same format as the default output
files from {\tt MultiDrizzle}, i.e., after correction for geometric distortion
using the WCS keywords in the image header. SExtractor (Bertin \& Arnouts 1996) 
was then run with a signal-to-noise threshold of S/N = 10 on each single drizzled
image. The resultant catalogs were trimmed on the basis of object
size and shape parameters, thereby rejecting the vast majority of cosmic
rays, CCD artifacts, and diffuse extended objects. Further catalog trimming
was done by visual inspection of the 2-dimensional structure of the catalog
sources, only retaining compact sources (non-saturated stars and compact
galaxies). Typically, 3\,--\,15 ``good'' sources remained available
for proper image alignment.  
Using these object catalogs, residual shifts (called 'delta shifts' in {\tt
  MultiDrizzle} nomenclature)  and rotations (${\Delta}x,~{\Delta}y,~{\Delta}
\theta$) between the individual singly drizzled images and a reference image
were determined using IRAF tasks {\tt xyxymatch} and {\tt geomap}. The
reference image was always taken to be the image observed first in the
visit. The resulting delta shifts and rotations were then fed to a second run
of {\tt MultiDrizzle} using {\tt driz\_separate = yes} to verify the result of
the alignment. This often prompted the need for a second iteration of the
alignment process. The formal uncertainties of the alignment process stayed
below 0.1 ACS/WFC pixel (as reported by the {\tt geomap} task). 

\subsection{Cosmic Ray Rejection}

Once the alignments between the individual images were determined, cosmic ray
rejection was done nominally in two steps. The first step was
performed by using {\tt MultiDrizzle} and its cosmic ray rejection
algorithm, which uses a process involving an image containing the
median values of each pixel in the (geometrically corrected and
aligned) input images as well as its derivative (in which the value of each 
pixel represents the largest gradient from the value of that pixel to
those of its direct neighbors; this image is used to avoid clipping
bright point sources) to simulate a ``clean'' version of the final
output image. For a typical 640-second F475W exposure, $\sim 
100,000 -  320,000$~pixels were affected by cosmic ray hits, i.e., $\sim
0.6\,--\,2$\% of the 4096$^2$ pixels. Thus, it is exceedingly rare for
a pixel to be affected by cosmic rays during all four exposures
(namely $\leq$\,3 pixels out of 4096$^2$). The number of pixels
affected by 3 cosmic rays out of 4 exposures was however significantly
higher (up to $\sim$\,120 pixels). In the latter cases, the median
value was replaced by the value of the one valid pixel if the median
value was larger than that of the valid pixel by a 5$\sigma$
threshold. This process yielded satisfactory results for the bulk of the
tiles, except for the central strip of pixels in each tile which only had
contributions from two exposures due to the gap between the two CCD
chips of ACS/WFC. Furthermore, the {\tt MultiDrizzle} output images of
the tiles which only had two or three successful 
exposures also contained several residual pixels affected by cosmic
rays. For the central strips in the tiles with four 
exposures and the full tiles with only two or three
exposures, we therefore used a second step of cosmic ray rejection by
means of the {\tt lacosmic} routine (van Dokkum 2001) using
parameter values chosen after careful testing. 

\subsection{Construction of Final ACS Images}

Before the final run of {\tt MultiDrizzle} on the input images to
produce the final image tiles using the cosmic ray masks
created as mentioned in the previous subsection, we created sky
variance maps for each individual exposure for the purpose of deriving
appropriate weight maps for the final image combination. These maps
contain all components of noise in the images, except for the Poisson
noise associated with the sources on the image. These maps were
constructed from the sky values determined by MultiDrizzle for each
individual exposure, the flatfield reference file, the dark current
reference file, and the read-out noise values as listed in the image
headers for each individual image quadrant (or read-out
amplifier). These sky variance images were then inverted and used as 
weight maps for the associated exposures. 

The final run of {\tt MultiDrizzle} was performed by shrinking the
input pixels by a factor 0.8 at the stage when the PSF is convolved by
the input pixel scale (i.e., the {\tt final\_pixfrac} parameter). This
factor was arrived at after some experimentation with different
values, and was a good match to the degree of subsampling induced by
the dither pattern we used. The final output images were produced in
the default unrotated frame of the ACS/WFC CCDs rather than with North
up so as to facilitate further analysis with PSF matching procedures.

\section{NICMOS parallel observations}

NICMOS parallel observations are made with the NIC3 camera of NICMOS in the 
J and H bands. The scientific motivations for these parallel observations, 
apart from that they are free, are the investigation of the near-IR luminosity 
function (LF), to compare with the SPITZER LF of Jenkins et al.\ (2007); a 
study of intra-cluster GCs,  where the combined ACS (g,I) and NICMOS (J,H) 
data can be used to separate intermediate-age GCs from old halo globulars 
(e.g., Puzia et al.\ 2002); and a study of the near-IR fundamental plane for 
dwarf galaxies, where the relative insensitivity of the IR luminosity to dust 
extinction reduces the scatter in in the FP.

The NIC3 camera has  a field of view of 51.2 arcsc square, with the
3 arcsecond dither to accomodate the ACS inter-CCD gap, the uniformly exposed
area is approximately 48 x 51 arcsec per visit. The NIC3 field centre is some 
8.5 arcminutes from the ACS field centre, at a position angle which depends
upon the telescope orentation. The NICMOS fields are thus fairly random 
positions in the cluster, and as can be seen from Figure \ref{fig:surveytiles}
only four of them overlap the area observed with ACS before the instrument
failure. NICMOS parallel observations 
were taken on all pre-SM4 visits, of which 21 have the full four dither 
positions, these are summarised in Table \ref{tab:NICMOS}.
\clearpage
\begin{deluxetable}{lllc}
\tablecolumns{4}
\tablewidth{0pc}
\tablecaption{\label{tab:NICMOS}NICMOS parallel observations - field centres and exposures}
\tablehead{
\colhead{Visit}&\colhead{RA}&\colhead{Dec}&\colhead{Exposure per filter (s)}}
\startdata
1&13:00:53.82&28:13:10.9& 2560 \\
2&13:00:38.21&28:13:10.9& 2560 \\
3&13:00:22.61&28:13:11.0& 1280 \\  
8&13:00:50.20&28:09:59.8& 2560 \\
9&13:00:34.61&28:09:59.9& 2560 \\
10&13:00:19.02&28:09:59.9& 2560 \\
12&12:59:47.81&28:09:59.8& 1920 \\
13&12:59:32.21&28:09:59.8& 1280 \\
14&12:59:16.61&28:09:59.9& 1920 \\
15&13:00:46.60&28:06:48.8& 2560 \\
16&13:00:31.01&28:06:48.9& 2560 \\
18&12:59:59.80&28:06:48.8& 2560 \\
19&12:59:44.21&28:06:49.9& 2560 \\
22&13:00:43.11&28:03:38.0& 2560 \\
23&13:00:27.40&28:03:38.0& 2560 \\
24&13:00:11.91&28:03:38.0& 2560 \\
25&12:59:56.31&28:03:38.0& 2560 \\
33&12:59:37.00&28:00:27.0& 2560 \\
45&12:58:49.90&27:33:21.2& 2560 \\
46&12:58:40.56&27:31:19.1& 2560 \\
55&12:57:22.73&27:38:58.6& 2560 \\
59&12:58:31.11&27:20:26.4& 2560 \\
63&12:56:26.40&27:21:58.4& 2560 \\
75&12:58:54.33&27:54:33.0& 2560 \\
78&12:57:37.50&27:30:20.6& 2560 \\
\enddata
\end{deluxetable}
\clearpage

In 2560 seconds exposure per filter the calculated 10$\sigma$ magnitude limits
for optimal extraction of point sources in the AB magnitude system are J = 25.9 
and H = 25.4.


\section{Data and Educational products}

Data products will include image data as processed by the ACS 
pipeline described above, and also object catalogs with a variety of 
structural parameters as described in Paper II and subsequent papers in this 
series.  

The calibrated ACS images produced by STScI have been ingested into
Astro-WISE\footnote{http://www.astro-wise.org/}(Valentijn et al 2006, Valentijn \& 
Verdoes Kleijn 2006). Astro-WISE connects 
the distributed research groups for the data
analysis and is used to create the source catalogs for the ACS images and to
model the surface brightness distributions of galaxies. The results will be
publicly available via the Astro-WISE web services. 

ACS data will be associated in Astro-WISE with all COMA Legacy Survey products 
including derived products and ancilliary data (e.g. Subaru, INT and UKIRT
imaging and multi-wavelength data). Astro-WISE also connects the survey 
to the EURO-VO.

Coma Legacy Survey data products will also be made available as part of the
MAST Treasury archive at 
STScI\footnote{http://archive.stsci.edu/hst/tall.html}. 
Expected timescales for the distribution of
data products are given in Table \ref{tab:DataProducts}.

An education and
public outreach (EPO) program has been designed whose goal is to share the 
valuable legacy dataset and the associated research with the public. The 
deliverables include:

\begin{enumerate}

\item
  $HST$ ACS images of Coma that will appear at StarDate Online Astronomy
  Picture of the Week;  in ViewSpace, which is seen in museums across the
  country; and in the revised  StarDate/Universo Teacher Guide;

\item
  An activity and DVD targeted at 9th-12th grade students; and revised Internet
  versions of the StarDate/Universo Teacher Guide, which will highlight research
  on  the Coma Cluster from $HST$ data.

\item
  Five Stardate radio programs on the Coma Cluster in English, Spanish and 
  German for distribution on monthly compact disk to over 500 radio stations 
  in the USA,  Mexico and Germany. 

\end{enumerate}

The EPO program is a coordinated effort between NASA, the University of Texas
at Austin, the McDonald Observatory Education and Outreach Office, and the Space
Telescope Science Institute (STScI)

\clearpage
\begin{deluxetable}{lll}
\tablecolumns{3}
\tablewidth{0pc}
\tablecaption{\label{tab:DataProducts}HST Data Products and Data Release Schedule}
\tablehead{
\colhead{Data Product}&\colhead{Description}&\colhead{Release date}}
\startdata
Raw Data & HST pipeline calibrated images& Immediate \\
Public catalog&Positions, magnitudes, geometry, color, & May 2008 \\
&\,\,\,\,from \textsc{SExtractor}& \\
Processed Image Data &Co-added, CR-cleaned, with best& December 2008 \\
&\,\,\,\, reference files& \\
Structural catalog& Multi-component structural analysis& December 2008 \\
External data & Redshifts and ground-based colors &May 2009\\
\enddata
\end{deluxetable}
\clearpage

\acknowledgments

This research and EPO program are supported by STScI through grants
HST-GO-10861 and HST-E0-10861.35-A. We acknowledge  Tony Roman and
Marco Chiaberge at STScI for their technical help.
S.J. acknowledges support from the National Aeronautics and Space
Administration (NASA) LTSA grant NAG5-13063, and NSF grant AST-0607748.
AWG is grateful for a Swinburne University of Technology RDS grant.
THP gratefully acknowledges support in form of a
Plaskett Fellowship at the Herzberg Institute of Astrophysics.
PE was supported by Deutsche Forschungsgemeinschaft Priority Program 1177.
RJS is supported under the STFC rolling grant PP/C501568/1
``Extragalactic Astronomy and Cosmology at Durham 2005--2010''.
R.G. acknowledges additional support from the NASA LTSA grant NAG5-11635.
BWM is supported by the Gemini Observatory, which is operated by the
Association of Universities for Research in Astronomy, Inc., on behalf
of the international Gemini partnership of Argentina, Australia, Brazil,
Canada, Chile, the United Kingdom, and the United States of America. We thank
the anonymous referee for very helpful comments.

\clearpage
\begin{deluxetable}{llllclll}
\tabletypesize{\scriptsize}
\rotate
\tablecolumns{8}
\tablewidth{0pc}
\tablecaption{\label{tab:Visits}List of survey fields}
\tablehead{
\colhead{Visit}&\colhead{Field}&\colhead{RA}&\colhead{Dec}&
\colhead{Orient}&\colhead{Members}&\colhead{Date of}&\colhead{Dither positions} \\
\colhead{}&\colhead{}&\multicolumn{2}{c}{(J2000.0)}&\colhead{}&\colhead{}&
\colhead{Observation}&\colhead{Obtained}}
\startdata
1&Coma1\_1&13:00:45.90&28:04:54.0&282.0&GMP2440,GMP2449,GMP2489& 09/Jan/2007 & 4 \\
2&Coma1\_2&13:00:30.30&28:04:54.0&282.0&GMP2626,GMP2559& 09/Jan/2007 & 4 \\
3&Coma1\_3&13:00:14.70&28:04:54.0&282.0&GMP2752,GMP2787,GMP2805,GMP2784,GMP2848,& 20/Jan/2007  & 2 \\
 &        &           &          &   &GMP2861,GMP2879,GMP2922 & \nodata & \nodata \\    
4&Coma1\_4&12:59:59.10&28:04:54.0&282&GMP3058& \nodata & \nodata \\
5&Coma1\_5&12:59:43.50&28:04:54.0&282&GMP3113,GMP3121,GMP3160& \nodata & \nodata \\
6&Coma7\_6&12:59:06.30&27:45:48.0&282&GMP3660,GMP3730,GMP3739,GMP3750& \nodata & \nodata \\
7&Coma1\_7&12:59:12.30&28:04:54.0&282&GMP3561,GMP3554,GMP3656& \nodata & \nodata \\
8&Coma2\_1&13:00:42.30&28:01:43.0&282.0&GMP2417,GMP2423,GMP2511,GMP2529,GMP2551,& 12/Jan/2007 & 4 \\
 &        &           &          &   &GMP2550,GMP2559& \nodata & \nodata \\
9&Coma2\_2&13:00:26.70&28:01:43.0&282.0&GMP2676,GMP2727,GMP2777& 13/Jan/2007 & 4 \\
10&Coma2\_3&13:00:11.10&28:01:43.0&282.0&GMP2839,GMP2856,GMP2940,GMP2960& 13/Jan/2007 & 4 \\
11&Coma2\_4&12:59:55.50&28:01:43.0&282&GMP3073& \nodata & \nodata \\
12&Coma2\_5&12:59:39.90&28:01:43.0&282.0&GMP3312& 12/Jan/2007 & 3 \\
13&Coma2\_6&12:59:24.30&28:01:43.0&282.0&GMP3390,GMP3406,GMP3433,GMP3438,GMP3471& 13/Jan/2007 & 2 \\
14&Coma2\_7&12:59:08.70&28:01:43.0&282.0&GMP3681,GMP3707,GMP3762,GMP3780,GMP3811& 08/Jan/2007 & 3 \\
15&Coma3\_1&13:00:38.70&27:58:32.0&282.0&GMP2510,GMP2516,GMP2535& 26/Jan/2007 & 4 \\
16&Coma3\_2&13:00:23.10&27:58:32.0&282.0&GMP2651,GMP2654,GMP2718,GMP2799,GMP2815,& 25/Jan/2007 & 4 \\
 &        &           &          &   &GMP2794,GMP2798& \nodata & \nodata \\
17&Coma3\_3&13:00:07.50&27:58:32.0&282&GMP2929,GMP2921,GMP2946,GMP2964,GMP2965,& \nodata & \nodata \\
 &        &           &          &   &GMP2985,GMP2940& \nodata & \nodata \\
18&Coma3\_4&12:59:51.90&27:58:32.0&282.0&GMP3018,GMP3098,GMP3146,GMP3166,GMP3170,& 25/Jan/2007 & 4 \\
 &        &           &          &   &GMP3206& \nodata & \nodata \\
19&Coma3\_5&12:59:36.30&27:58:32.0&282.0&GMP3206,GMP3213,GMP3254,GMP3269,GMP3291,& 25/Jan/2007 & 4 \\
 &        &           &          &   &GMP3292,GMP3308,GMP3329,GMP3367,GMP3414& \nodata & \nodata \\
20&Coma3\_6&12:59:20.70&27:58:32.0&282&GMP3471,GMP3484,GMP3487,GMP3489,GMP3515,& \nodata & \nodata \\
 &        &           &          &   &GMP3534,GMP3565,GMP3602,GMP3615,GMP3639& \nodata & \nodata \\
21&Coma3\_7&12:59:05.10&27:58:32.0&282&GMP3664,GMP3761,GMP3851,GMP3877,GMP3794& \nodata & \nodata \\
22&Coma4\_1&13:00:35.10&27:55:21.0&282.0&GMP2541,GMP2563,GMP2571,GMP2585,GMP2591& 22/Jan/2007 & 4 \\
23&Coma4\_2&13:00:19.50&27:55:21.0&282.0&GMP2692,GMP2736,GMP2780,GMP2778& 27/Jan/2007 & 4 \\
24&Coma4\_3&13:00:03.90&27:55:21.0&282.0&GMP2931,GMP3017,GMP3034& 24/Jan/2007 & 4 \\
25&Coma4\_4&12:59:48.30&27:55:21.0&282.0&GMP3068,GMP3080,GMP3133,GMP3131,GMP3201,& 24/Jan/2007 & 4 \\
 &        &           &          &   &GMP3215,GMP3222& \nodata & \nodata \\
26&Coma4\_5&12:59:32.70&27:55:21.0&282&GMP3296,K9992,GMP3325,GMP3340,GMP3352,& \nodata & \nodata \\
 &        &           &          &   &GMP3376,GMP3424& \nodata & \nodata \\
27&Coma4\_6&12:59:17.10&27:55:21.0&282&GMP3486,GMP3510,GMP3511,GMP3522,GMP3645& \nodata & \nodata \\
28&Coma4\_7&12:59:01.50&27:55:21.0&282&GMP3719,GMP3782,GMP3855& \nodata & \nodata \\
29&Coma5\_1&13:00:31.50&27:52:10.0&282&GMP2716& \nodata & \nodata \\
30&Coma5\_2&13:00:15.90&27:52:10.0&282&GMP2716,GMP2753,GMP2852,GMP2897,GMP2910,& \nodata & \nodata \\
 &        &           &          &   &GMP2913& \nodata & \nodata \\
31&Coma5\_3&13:00:00.30&27:52:10.0&282&GMP2910,GMP3052& \nodata & \nodata \\
32&Coma5\_4&12:59:44.70&27:52:10.0&282&GMP3178,GMP3196,GMP3205& \nodata & \nodata \\
33&Coma5\_5&12:59:29.10&27:52:10.0&282.0&GMP3311,GMP3339,GMP3383,GMP3400,GMP3423,& 15/Jan/2007 & 4 \\
 &        &           &          &   &GMP3473& \nodata & \nodata \\
34&Coma5\_6&12:59:13.50&27:52:10.0&282&GMP3557,GMP3645,GMP3706,GMP3719,GMP3733& \nodata & \nodata \\
35&Coma5\_7&12:58:57.90&27:52:10.0&282&GMP3821,GMP3911,GMP3946& \nodata & \nodata \\
36&Coma6\_1&13:00:27.90&27:48:59.0&282&GMP2603& \nodata & \nodata \\
37&Coma6\_2&13:00:12.30&27:48:59.0&282&GMP2783,GMP2800,GMP2956& \nodata & \nodata \\
38&Coma6\_3&12:59:56.70&27:48:59.0&282&GMP3092,GMP3122,GMP3126& \nodata & \nodata \\
39&Coma6\_4&12:59:41.10&27:48:59.0&282&GMP3126,GMP3313,GMP3324& \nodata & \nodata \\
40&Coma6\_5&12:59:25.50&27:48:59.0&282&GMP3403,GMP3411,GMP3425& \nodata & \nodata \\
41&Coma6\_6&12:59:09.90&27:48:59.0&282& \nodata & \nodata & \nodata \\
42&Coma6\_7&12:58:54.30&27:48:59.0&282&GMP3895,GMP3896,GMP3898,GMP3925,GMP3943,& \nodata & \nodata \\
 &        &           &          &   &GMP3949,GMP3997& \nodata & \nodata \\
43&Outskirts\_1&12:57:56.90&27:28:55.0&239-241&GMP4522,GMP4597,GMP4630& \nodata & \nodata \\
44&Outskirts\_2&12:58:01.70&27:25:48.0&265-270&GMP4479,GMP4535,GMP4577,GMP4568& \nodata & \nodata \\
45&Outskirts\_3&12:58:21.58&27:27:40.7&318.0&GMP4206,GMP4330,GMP4381& 21/Nov/2006 & 4 \\
46&Outskirts\_4&12:58:34.00&27:22:58.8&280.0&GMP4192,GMP4215& 17/Jan/2007 & 4 \\
47&Outskirts\_5&12:57:27.38&27:28:59.1&219-221&GMP4792,GMP4794,GMP4928,GMP4943,GMP4956& \nodata & \nodata \\
48&Outskirts\_6&12:58:00.50&27:22:23.7&238-290&GMP4484,GMP4545,GMP4565,GMP4578& \nodata & \nodata \\
49&Outskirts\_7&12:57:42.52&27:26:37.6&280-323&GMP4692,GMP4712,GMP4768& \nodata & \nodata \\
50&Outskirts\_8&12:57:27.40&27:23:37.0&230-323&GMP4910,GMP4918& \nodata & \nodata \\
51&Outskirts\_9&12:57:04.70&27:21:25.1&260-323&GMP5076,GMP5136& \nodata & \nodata \\
52&Outskirts\_10&12:56:44.30&27:26:12.0&250-323&GMP5250,GMP5255,GMP5296& \nodata & \nodata \\
53&Outskirts\_11&12:57:55.25&27:13:55.7&any&GMP4591& \nodata & \nodata \\
54&Outskirts\_12&12:56:37.81&27:33:08.4&215-255&GMP5284,GMP5320,GMP5364& \nodata & \nodata \\
55&Outskirts\_13&12:57:04.22&27:31:34.5&299.1&GMP5102& 19/Jan/2007 & 4 \\
56&Outskirts\_14&12:57:26.40&27:32:54.4&220-323&GMP4852,GMP4907,GMP4937& \nodata & \nodata \\
57&Outskirts\_15&12:58:02.08&27:33:54.0&250-260&GMP4447,GMP4469,GMP4579& \nodata & \nodata \\
58&Outskirts\_16&12:58:34.59&27:33:43.3&220-225&GMP4117,GMP4156,GMP4255& \nodata & \nodata \\
59&Outskirts\_17&12:58:31.15&27:11:58.5&270.05&GMP4135,GMP4259,GMP4294& 16/Jan/2007 & 4 \\
60&Outskirts\_18&12:58:15.55&27:05:15.1&any&GMP4383& \nodata & \nodata \\
61&Outskirts\_19&12:58:48.72&28:00:52.8&265-323&GMP3969,GMP3973,GMP4003& \nodata & \nodata \\
62&Outskirts\_20&12:56:36.03&26:54:17.8&any&GMP5361& \nodata & \nodata \\
63&Outskirts\_21&12:56:29.80&27:13:32.8&265.0&GMP5335,GMP5359,GMP5365,GMP5472& 21/Jan/2007 & 4 \\
64&Outskirts\_22&12:57:56.51&27:02:16.4&any&GMP4582,GMP4596& \nodata & \nodata \\
65&Outskirts\_23&12:57:21.08&27:00:04.2&220-260&GMP4961,GMP4980,GMP4947& \nodata & \nodata \\
66&Outskirts\_24&12:57:08.88&27:05:47.8&200-290&GMP5032,GMP5097& \nodata & \nodata \\
67&Outskirts\_25&12:56:37.84&27:02:49.3&230-250&GMP5283,GMP5293,GMP5294,GMP5395& \nodata & \nodata \\
68&Outskirts\_26&12:57:14.03&27:15:14.7&any&GMP5012,GMP5019& \nodata & \nodata \\
69&Outskirts\_27&12:59:12.21&27:36:56.9&265&GMP3585,GMP3598,GMP3696& \nodata & \nodata \\
70&Outskirts\_28&12:57:21.38&27:37:42.1&240-250&GMP4888,GMP4905,GMP4967,GMP4987& \nodata & \nodata \\
71&Outskirts\_29&12:56:29.63&27:42:27.2&250-295&GMP5345,GMP5362,GMP5434& \nodata & \nodata \\
72&Outskirts\_30&12:55:38.57&27:44:39.9&284-286&GMP5850,GMP5912& \nodata & \nodata \\
73&Outskirts\_31&12:57:03.95&27:45:06.4&225-235&GMP5096,GMP5100& \nodata & \nodata \\
74&Outskirts\_32&12:58:25.87&26:54:23.2&245-255&GMP4183,GMP4340& \nodata & \nodata \\
75&Coma7\_7&12:58:47.75&27:46:12.7&280.0&GMP3949,GMP3958,GMP4035,GMP4060& 17/Jan/2007 & 4 \\
76&Outskirts\_34&12:58:19.13&27:45:44.6&any&GMP4341,GMP4364& \nodata & \nodata \\
77&Outskirts\_35&12:58:03.52&27:40:57.8&any&GMP4502,GMP4557& \nodata & \nodata \\
78&Outskirts\_36&12:57:10.80&27:24:18.0&314.52&GMP5038& 27/Nov/2006 & 4 \\
79&Outskirts\_37&12:57:50.52&27:38:38.8&any&GMP4632& \nodata & \nodata \\
80&Outskirts\_38&12:56:04.00&27:09:03.0&any&GMP5676& \nodata & \nodata \\
81&Outskirts\_39&12:59:50.50&27:44:48.9&230&GMP3071,GMP3132,GMP3176,GMP3192,GMP3195,& \nodata & \nodata \\
 &        &           &          &   &GMP3211& \nodata & \nodata \\
82&Outskirts\_40&12:56:14.60&27:30:23.2&any&GMP5546& \nodata & \nodata \\ \hline
\enddata
\end{deluxetable}


\begin{thebibliography}{}
\bibitem{ANS06} Abadi, M.G., Navarro, J.F., \& Steinmetz, M.\ 2006, \mnras, 365, 747
\bibitem[Adami et al. (2006)]{Ad06} Adami, C. et al. 2006, \aap, 451, 1159
\bibitem[Adelman-McCarthy et al. (2007)]{SDSSDR5} Adelman-McCarthy, J.K.
et al. 2007, \apjs, in press.
\bibitem[Aguerri et al. (2001)]{Agu01} Aguerri, J.A.L., Balcells, M. \&
Peletier, R.F. 2001, \aap, 367, 428
\bibitem[Aguerri et al. (2005)]{Agu05} Aguerri, J.A.L., Iglesias-P\'aramo, J.,
V\'ilchez, J.M., Mu\~noz-Tu\~n\'on, C. \& S\'anchez-Janssen, R. 2005, \aj, 130, 475
\bibitem[Ashman \& Zepf (1998)]{az98} Ashman, K.~M., \& Zepf, 
S.~E.\ 1998, Globular cluster systems, Cambridge University Press
\bibitem[Athanassoula (2005)]{Ath05} Athanassoula, E. 2005 \mnras, 358, 1477
\bibitem[Babul \& Rees (1992)]{Bab92} Babul, A. \& Rees, M.J. 1992,
\mnras, 255, 346
\bibitem[Bai et al. (2006)]{Bai06} Bai L., Rieke, G.H., Rieke, M.J., 
Hinz, J.L., Kelly, D.M. \& Blaylock, M. 2006, \apj, 639, 827
\bibitem[Balcells et al. (2003)]{Bal03} Balcells, M., Graham, A.W., 
Dom\'inguez-Palmero, L. \& Peletier, R.F. 2003, \apjl, 582, L79
\bibitem{BGP07} Balcells, M., Graham, A.W., \& Peletier, R.F.\ 2007, \apj, in press
\bibitem[Barnes \& Hernquist (1992)]{BaHe92} Barnes, J. \& Hernquist, L.
1992, \araa, 30, 705
\bibitem{Bet6a} Bassino, L.P., Faifer, F.R., Forte, J.C., Dirsch, B., Richtler, T., 
Geisler, D. \& Schuberth, Y. 2006, \aap, 451, 789 
\bibitem[Baum et al. (1995)]{Baum95} Baum, W.A. et al. 1995, \aj, 110, 2537
\bibitem[Baum et al. (1997)]{Baum97} Baum, W.A., Hammergren, M.,
Thomsen, B., Groth, E.J., Faber, S.M., Grillmair, C.J. \& Ajhar, E.A.
 1997, \aj, 113, 1483
\bibitem{BBR80} Begelman, M.C., Blandford, R.D., \& Rees, M.J.\ 1980, Nature, 287, 307
\bibitem[Bekki \& Shioya (1999)]{Bek99} Bekki, K., \& Shioya, Y. 1999, \apj, 513, 108
\bibitem[Bekki et al. (2001)]{Bek01} Bekki, K., Couch, W.J., \& Drinkwater, M. 2001,
\apjl, 552, L105
\bibitem{BaY06} Bekki, K. \& Yahagi, H.\ 2006, \mnras, 372, 1019 
\bibitem[Bender et al. (1992)]{Ben93} Bender, R., Burstein, D. \& Faber, S.M.
1992, \apj 399, 462
\bibitem[Bernstein et al. (1995)]{Ber95} Bernstein, G.M., Nichol, R.C., 
Tyson, J.A., Ulmer, M.P. \& Wittman, D. 1995, \aj, 110, 1507
\bibitem[Bertin \& Arnouts (1996)]{BA96} Bertin, E. \& Arnouts, S. 1996,
\aaps, 117, 393
\bibitem{BBJ00} Binggeli, B., Barazza, F., \& Jerjen, H.\ 2000, \aap, 359, 447 
\bibitem[Blanton et al. (2005)]{Bla05} Blanton, M.R., Lupton, R.H., 
Schlegel, D.J., Strauss, M.A., Brinkmann, J., Fukugita, M. \& Loveday, J.
2005, \apj, 631, 208
\bibitem[Bonamente et al. (2003)]{Bon03} Bonamente, M., Joy, M.K. \& Lieu, R.
2003, \apj 585, 722.
\bibitem[Bothun et al. (1984)]{Bot84} Bothun, G.D., Schommer, R.A. \&
Sullivan, W.T. 1984, \aj, 89, 466
\bibitem[Bower et al. (1992a)]{Bow92a} Bower, R.G., Lucey, J.R. \& Ellis, R.S.
1992a, \mnras, 254, 589
\bibitem[Bower et al. (1992b)]{Bow92b} Bower, R.G., Lucey, J.R. \& Ellis, R.S.
1992b, \mnras, 254, 601
\bibitem[Bowyer et al. (2004)]{Bowy04} Bowyer, S., Korpela, E.J., Lampton, M.
\& Jones, T.W. 2004, \apj 605, 168
\bibitem[Bravo-Alfuro et al. (2000)]{BrA00} Bravo-Alfuro, H., Cayatte, V.,
van Gorkom, J.H. \& Balkowski, C. 2000, \aj, 119, 580
\bibitem[Bravo-Alfuro et al. (2001)]{BrA01} Bravo-Alfuro, H., Cayatte, V.,
van Gorkom, J.H. \& Balkowski, C. 2001, \aap, 379, 347 
\bibitem[Briel et al. (2001)]{Bri01} Briel, U.G., et al. 2001, \aap, 365, L60
\bibitem{BaO84} Butcher, H., \& Oemler, A., Jr.\ 1984, \apj, 285, 426 
\bibitem[Caldwell (1983)]{Cal83} Caldwell, N. 1983, \aj, 88, 804
\bibitem[Caldwell et al. (1993)]{Cal93} Caldwell, N., Rose, J.A., Sharples, 
R.M., Ellis, R.S. \& Bower, R.G. 1993, \aj, 106, 473
\bibitem[Carlberg (1984)]{Car84} Carlberg, R.G. 1984, \apj, 286, 416
\bibitem[Caon et al. (1993)]{Cao93} Caon, N., Capaccioli, M. \& D'Onofrio, M.
1993, \mnras, 265, 1013
\bibitem[Cody et al. (2007)]{Cod07} Cody, A.M., Carter, D. Bridges, T.J.,
Mobasher, B. \& Poggianti, B.M. 2007, \mnras, submitted.
\bibitem[Cole et al. (2000)]{Col00} Cole, S., Lacey, C.G., Baugh, C.M., \&
Frenk, C.S. 2000, \mnras, 319, 168
\bibitem[Colless \& Dunn (1996)]{CD96} Colless, M. \& Dunn, A.M. 1996,
\apj 458, 435
\bibitem{Con03b} Conselice, C.~J., O'Neil, K., Gallagher, J.~S., \& Wyse, R.~F.~G.\ 
2003, \apj, 591, 167 
\bibitem[Cooray \& Cen (2005)]{Coo05} Cooray, A. \& Cen, R. 2005, \apjl, 633, 
L69
\bibitem{Cote6} Cot\'e, P., et al.\ 2006, \apjs, 165, 57
\bibitem{Det06} Debattista, V.P., Ferreras, I., Pasquali, A., Seth, A., 
De Rijcke, S. \& Morelli, L. 2006, \apj, 651, L97
\bibitem[De Propris et al. (2005)]{DeP05} De Propris, R., Phillipps, S., 
Drinkwater, M.J., Gregg, M.D., Jones, J.B., Evstigneeva, E. \& Bekki, K.
2005, \apj, 623, 105
\bibitem[Dressler (1980)]{Dre80} Dressler, A.R. 1980, \apj, 236, 351
\bibitem[Drinkwater et al (2000)]{Drin00} Drinkwater, M.J., et al. 2000, 
\pasa, 17, 227
\bibitem[Drinkwater et al (2003)]{Drin03} Drinkwater, M.J., et al. 2003, 
Nature, 423, 519
\bibitem[Driver et al (1998)]{Dri98} Driver, S.P., Couch, W.J. \& Phillipps, S.
1998, \mnras, 301, 369
\bibitem{Dri7a}Driver, S.~P., Allen, P.~D., Liske, J., \& Graham, 
A.~W.\ 2007a, \apj, 657, L85
\bibitem{Dri7b}Driver, S.P., Popescu, C.C., Tuffs, R.J., Liske, J., 
Graham, A.W., Allen, P.D., De Propris, R.\ 2007b, \mnras, in press 
(arXiv:0704.2140)
\bibitem[Duc et al. (2004)]{Duc04} Duc, P.-A., Bournard, F. \& Masset, F.
2004, \aap 427, 803
\bibitem{EMO91} Ebisuzaki, T., Makino, J., \& Okumura, S.K.\ 1991, Nature, 354, 212
\bibitem[Edwards et al. (2002)]{Edw02} Edwards, S.A., Colless, M., 
Bridges, T.J., Carter, D., Mobasher, B. \& Poggianti, B.M. 2002,
\apj, 567, 178	
\bibitem[Efstathiou (2000)]{Efs00} Efstathiou, G. 2000, \mnras, 317, 697
\bibitem{Ein65} Einasto, J.\ 1965, Trudy Inst.\ Astrofiz.\ Alma-Ata, 5, 87
\bibitem{Ein72} Einasto, J.\ 1972, Tartu, Academy of Sciences of the Estonian SSR, 
Institute of Physics and Astronomy, W.~Struve Astrophysical Observatory, 1972., 13 
\bibitem{Ein74} Einasto, J.\ 1974, Stars and the Milky Way System, 291 
\bibitem[Eisenhardt et al. (2007]{Eis07} Eisenhardt, P.R., De Propris, R., 
Gonzalez, A.H., Stanford, S.A., Wang, M., \& Dickinson, M. 2007,
\apjs, 169, 225
\bibitem[Elmegreen et al. (1990)]{Elm90} Elmegreen, D.~M., Elmegreen, B.~G.,
\& Bellin, A.~D.\ 1990, \apj, 364, 415
\bibitem[Elmegreen et al. (2004)]{Elm04} Elmegreen, B.~G., Elmegreen, D.~M., 
\& Hirst, A.~C.\ 2004, \apj, 612, 191
\bibitem[Erwin \& Sparke(2002)]{erwin02} Erwin, P., \& Sparke, L. S. 2002,
\aj, 124, 65
\bibitem[Erwin(2005)]{erwin05} Erwin, P. 2005, \mnras, 364, 283
\bibitem[Erwin et al.(2003)]{erwin03} Erwin, P., Vega Beltr\'an, J. C.,
Graham, A. W., \& Beckman, J. E., \& Pohlen, M. 2003, \apj, 597, 929
\bibitem[Erwin et al.(2005)]{erwin05-apjl} Erwin, P., Beckman, J. E.,
\& Pohlen, M. 2005, \apjl, 626, L81
\bibitem[Erwin et al.(2007)]{erwin07} Erwin, P., Pohlen, M., \& Beckman, J. E. 2007,
\aj, in press
\bibitem[Faber \& Jackson (1976)] {FJ76} Faber, S.M. \& Jackson, R.E. 1976,
\apj, 204, 668
\bibitem[Fellhauer \& Kroupa (2002)]{FK02} Fellhauer, M. \& Kroupa, P. 2002,
\mnras, 330, 642
\bibitem[Fellhauer \& Kroupa (2005)]{FK05} Fellhauer, M. \& Kroupa, P. 2005,
\mnras, 359, 223
\bibitem{FaB84} Ferguson, H., \& Binggeli, B.\ 1994, \araa, 6, 67
\bibitem{FaS89} Ferguson, H., \& Sandage, A.\ 1989 \apj, 346, L53
\bibitem[Ferguson \& Sandage (1991)]{Fer91} Ferguson, H.C. \& Sandage, A.R.
1991, \aj, 101, 765
\bibitem{FaM00} Ferrarese, L., \& Merritt, D.\ 2000, \apj, 539, L9
\bibitem{Fet6a} Ferrarese, L., et al.\ 2006a, \apjs, 164, 334
\bibitem{Fet6b} Ferrarese, L., et al.\ 2006b, \apj, 644, L21
\bibitem[Finoguenov et al. (2004)]{Fin04} Finoguenov, A., Briel, U.G., 
Henry, J.P., Gavazzi, G., Iglesias-Paramo, J. \& Boselli, A. 2004, \aap, 419, 
47
\bibitem[Forbes et al. (2005)]{For05} Forbes, D., S\'anchez-Bl\'azquez, P.,
\& Proctor, R. 2005, \mnras, 361, 6
\bibitem[Freeman(1970)]{freeman70} Freeman, K. C. 1970, \apj, 160, 811
\bibitem[Fruchter \& Hook (2002)]{fruhoo02} Fruchter, A. S., \& Hook,
  R. N.\ 2002, \pasp, 114, 144
\bibitem[Gardner et al. (1997)]{Gar97} Gardner, J.P., Sharples, R.M., Frenk,
C.S. \& Carrasco, B.E. 1997, \apjl, 480, L99
\bibitem[Gavazzi et al. (2006)]{Gav06} Gavazzi, G., O'Neil, K., Boselli, A.
\& van Driel, W. 2006, \aap 449, 929
\bibitem{Get00} Gebhardt, K., et al.\ 2000, \aj, 119, 1157
\bibitem[Geller \& Huchra (1989)]{GH89} Geller, M.J. \& Huchra, J.P. 1989,
Science, 246, 897
\bibitem[Godwin \& Peach (1977)]{GP77} Godwin, J.G. \& Peach, J.V. 1977,
\mnras, 181, 323
\bibitem[Godwin et al. (1983)]{GMP83} Godwin, J.G., Metcalfe, N. \& Peach, J.V.
1983, \mnras, 202, 113
\bibitem{Gra04} Graham, A.W.\ 2004, \apj, 613, L33
\bibitem{Gra07} Graham, A.W.\ 2007, MNRAS, in press
\bibitem{GaD07} Graham, A.W., \& Driver, S.P.\ 2007, \apj, 655, 77
\bibitem{GECT1} Graham, A.W., Erwin, P. Caon, N., \& Trujillo, I.\ 2001, 
\apj, 563, L11
\bibitem{GETAR} Graham, A.W., Erwin, P., Trujillo, I., \& Asensio-Ramos, A.\ 
2003, \aj, 125, 2951
\bibitem{GaG03} Graham, A.W., \& Guzm\'an, R.\ 2003, \aj, 125, 2936
\bibitem{GJG03} Graham, A.W., Jerjen, H., \& Guzm\'an, R.\ 2003, \aj, 126, 1787
\bibitem[Gregg (1997)]{Gre97} Gregg, M.D. 1997 New Astronomy, 1, 363
\bibitem[Gualandris \& Merritt (2007)]{Gua} Gualandris, A. \& Merritt D. 
2007, \apj, submitted
(arXiv:0708.0771)
\bibitem[Guti\'errez, C.M. et al. (2004)]{Gut04} Guti\'errez, C.M., Trujillo, 
I., Aguerri, J.A.L., Graham, A.W. \& Caon, N. 2004, \apj, 602, 664
\bibitem[Guzm\'an et al. (1992)]{Guz92} Guzm\'an. R., Lucey, J.R., Carter, D. \& 
Terlevich, R.J. 1992, \mnras, 257, 187
\bibitem[Guzm\'an et al. (1993)]{Guz93} Guzm\'an. R., Lucey, J.R. \& Bower, R.G. 1993,
\mnras, 265, 731
\bibitem[Hammer et al (2008)]{Ham08} Hammer, D. et al. 2008, in preparation.
\bibitem[Harris et al. (2000)]{Har00} Harris, W.E., Kavelaars, J.J.,
Hanes, D.A., Hesser, J.E. \& Pritchet, C.J. 2000, \apj, 533, 137
\bibitem{HaR04} H\"aring, N., \& Rix, H.-W.\ 2004, \apj, 604, L89
\bibitem[Ha\c{s}egan et al. (2005)]{Ha05} Ha\c{s}egan, M. et al. 2005,
\apj, 627, 203
\bibitem[Hilker(2006)]{Hil06} Hilker, M.\ 2006, "Globular Clusters - Guides to Galaxies", 
Concepcion, Chile, eds. T. Richtler \& S. Larsen (Springer), see also arXiv:astro-ph/0605447
\bibitem[Hilker et al.(1999)]{Hil99} Hilker, M., Infante, L., 
Vieira, G., Kissler-Patig, M., \& Richtler, T.\ 1999, \aaps, 134, 75
\bibitem[Hilker et al. (2004)]{Hil04} Hilker, M., Kayser, A., Richtler, T.
\& Willemsen, P. 2004, \aap, 422, L9
\bibitem[Hjorth \& Tanvir (1997)]{Hjo97} Hjorth, J. \& Tanvir, N.R. 1997, \apj, 482, 68
\bibitem[Hornschemeier et al. (2006)]{Hor06} Hornschemeier, A.E., Mobasher, B.,
Alexander, D.M., Bauer, F.E., Bautz, M.W., Hammer, D. \& Poggianti, B.M.
2006, \apj, 643, 144
\bibitem[James et al. (2006)] {Jam06} James, P.A., Salaris, M., Davies, J.I.,
Phillipps, S. \& Cassisi, S. 2006, \mnras, 367, 339
\bibitem[Jenkins et al. (2007)]{Jen07} Jenkins, L.P., Hornschemeier, A.E., 
Mobasher, B., Alexander, D.M. \& Bauer, F. E. 2007 \apj, in press (arXiv:0705.3681v1 [astro-ph])
\bibitem[Jensen et al. (1999)]{Jens99} Jensen, J.B., Tonry, J. \& Luppino, G.
1999 \apj, 510, 71
\bibitem[Jogee (1999)]{Jog99} Jogee, S.  1999, Ph.D. thesis, Yale University
\bibitem[Jogee et al. (2004)]{Jog04} Jogee, S. et. al. 2004, \apjl, 615, L105
\bibitem[Jogee et al. (2005)]{Jog05} Jogee, S., Scoville, N., \& 
Kenney, J.~D.~P.\ 2005, \apj, 630, 837
\bibitem[Jones et al. (2006)]{Jon06} Jones, J.B. et al. 2006, \aj, 131, 312
\bibitem[Jord\'an et al. (2005)]{Jord05} Jord\'an, A. et al. 2005, \apj, 634,
1002
\bibitem{Jet07} Jord\'an, A., et al.\ 2007, \apjs, 169, 213
\bibitem[Jorgensen (1999)]{Jorg99} Jorgensen, I. 1999, \mnras, 306, 607
\bibitem[Jorgensen et al. (1996)]{Jorg96} Jorgensen, I., Franx, M. \&
Kjaergaard, P. 1996, \mnras, 280, 167
\bibitem[Jorgensen et al. (2006)]{Jor06} Jorgensen, I., Chiboucas, K., 
Flint, K., Bergmann, M., Barr, J. \& Davies, R.L. 2006, \apjl 639, L9
\bibitem[Kaastra et al. (2003)]{Kaa03} Kaastra, J.S., Lieu, R., Tamura, T., 
Paerels, F.B.S. \& den Herder, J. W. 2003, \aap 397, 445
\bibitem[Kavelaars et al. (2000)]{Kav00} Kavelaars, J.J., Harris, W.E.,
Hanes, D.A., Hesser, J.E. \& Pritchet, C.J. 2000, \apj, 533, 125
\bibitem[Knapen(2005)]{knapen05} Knapen, J. H. 2005, \aap, 429, 141
\bibitem[Kobayashi (2004)]{Kob04} Kobayashi, C. 2004, \mnras, 347, 740
\bibitem[Kochanek et al. (2001)]{Koc01} Kochanek, C.S. et al. 2001, \apj, 
560, 566
\bibitem[Koekemoer et al. (2003)]{koek+03} Koekemoer, A. M., Fruchter,
  A. S., Hook, R. N., \& Hack, W.\ 2003, in ``2002 HST
  Calibration Workshop'', eds.\ S. Arribas, A. Koekemoer, \& B. Whitmore
  (Baltimore: STScI), 337
\bibitem[Komiyama et al. (2002)]{Kom02} Komiyama, Y. et al. 2002,
\apjs, 138, 265
\bibitem[Kormendy (1985)]{Kor85} Kormendy, J. 1985, \apj, 295, 73
\bibitem{Kor88} Kormendy, J.\ 1988, \apj, 325, 128
\bibitem[Kormendy \& Kennicutt (2004)]{KK04} Kormendy, J. \& Kennicutt, R.C.
2004, \araa, 42, 603
\bibitem[Laine et al.(2002)]{laine02} Laine, S., Shlosman, I., Knapen, J. H.,
\& Peletier, R. F. 2002, \apj, 567, 97
\bibitem[Laurikainen et al. (2005)]{Lau05} Laurikainen, E., Salo, H., 
\& Buta, R.\ 2005, \mnras, 362, 1319
\bibitem[Lieu et al, (1996)]{Lie96} Lieu, R., Mittaz, J.P.D., Bowyer, S., 
Breen, J.O., Lockman, F.J., Murphy, E.M. \& Hwang, C.-Y.  1996, Science, 
274, 1335
\bibitem[Lisker et al.(2006)]{lisker06} Lisker, T., Debattista, V. P.,
Ferreras, I., \& Erwin, P. 2006, \mnras, 370, 477
\bibitem[Lisker et al. (2007)]{Lis07} Lisker, T., Grebel. E.K., Bingelli, B.
\& Glatt, K. 2007, \apj, 660, 1186
\bibitem[Lucas et al. (2006)]{lucas+06} Lucas, R. A., Swam, M.,
  Mutchler, M., \& Sirianni, M.\ 2006, in ``2005 HST
  Calibration Workshop'', eds.\ A. Koekemoer, P. Goudfrooij, \&
  L. Dressel (Baltimore: STScI), 61  
\bibitem[Lucey et al. (1991)]{Luc91} Lucey, J.R., Guzm\'an, R., Carter, D. \&
Terlevich, R.J. 1991, \mnras, 253, 584
\bibitem[Marconi \& Hunt (2003)]{MaH03} Marconi, A., \& Hunt, L.K.\ 2003, \apj, 589, L21
\bibitem[Mar\'in-Franch \& Aparicio (2002)]{MFA02} Mar\'in-Franch, A. \&
Aparicio, A. 2002, \apj, 568, 174
\bibitem[Mastropietro et al. (2005)]{Mas05} Mastropietro, C, Moore, B., 
Mayer, L., Debattista, V. P., Piffaretti, R. \& Stadel, J. 2005, \mnras, 364, 607
\bibitem[Matkovi\'c \& Guzm\'an (2005)]{MG05} Matkovi\'c, A. \& Guzm\'an, R.
2005, \mnras 362, 289
\bibitem[Matkovi\'c et al. (2007)]{Mat07} Matkovi\'c, A., Guzm\'an, R.,
S\'anchez-Bl\'azquez, P., Gorgas, J., Cardiel, N. \& Gruel, N. 2007, \mnras submitted
\bibitem[Matkovi\'c \& Guzm\'an (2008)]{MG08} Matkovi\'c, A. \& Guzm\'an, R.
2008, in preparation
\bibitem{MaD02} McLure, R.J., \& Dunlop, J.S.\ 2002, \mnras, 331, 795
\bibitem[Mehlert et al. (2003)]{Meh03} Mehlert, D., Thomas, D., Saglia, R.P.,
Bender, R. \& Wegner, G. 2003, \aap, 407, 423
\bibitem{Mer84} Merritt, D.\ 1984, \apj, 280, L5
\bibitem{Mer6a} Merritt, D.\ 2006a, Reports of Progress in Physics, 69, 2513.
\bibitem{Mer6b} Merritt, D.\ 2006b, \apj, 648, 976
\bibitem[Merritt et al.(2004)]{2004ApJ...607L...9M} Merritt, D.,
Milosavljevi{\'c}, M., Favata, M., Hughes, S.~A., \& Holz, D.~E.\ 2004,
\apjl, 607, L9 
\bibitem{MGMDT} Merritt, D., Graham, A.W., Moore, B., Diemand, J., \& Terzic, B.\ 2006, \aj, 132, 2685
\bibitem[Merritt et al.(2007)]{2007arXiv0705.2745M} Merritt, D., Mikkola,
S., \& Szell, A.\ 2007, ArXiv e-prints, 705, arXiv:0705.2745
\bibitem[Mieske et al. (2006)]{Mie06} Mieske, S. et al. 2006, \apj, 653, 193
\bibitem[Mieske et al. (2007)]{Mie07} Mieske, S., Hilker, M., Infante, L. \&
Mendes de Oliveira, C. 2007, in ``Dark Galaxies and lost Baryons'', Proc.
I.A.U. symposium 244, eds: J.I. Davies \& M.J. Disney, p119, (astro-ph/0707.3869)
\bibitem[Milne et al. (2007)]{Mil07} Milne, M.L., Pritchet, C.J., Poole, G.B.,
Gwyn, S.D.J., Kavelaars, J.J., Harris, W.E. \& Hanes, D.A. 2007, \aj, 133, 177
\bibitem[Mobasher et al. (2001)]{Mob01} Mobasher, B. et al. 2001, \apjs, 137, 279
\bibitem[Mobasher et al. (2003)]{Mob03} Mobasher, B. et al. 2003, \apj, 587, 605
\bibitem{Met06} Moore, B., et al.\ 2006, \mnras, 368, 563
\bibitem{Moo96} Moore, B., Katz, N., Lake, G., Dressler, A., \& 
Oemler, A.\ 1996, \nat, 379, 613 
\bibitem[Moore et al. (1998)]{Mo98} Moore, B., Lake, G. \& Katz, N. 1998,
\apj, 495, 139
\bibitem[Moore et al. (2002)]{Mo02} Moore, S.A.W., Lucey, J.R., Kuntschner, H.
\& Colless, M. 2002, \mnras, 336, 382
\bibitem[Navarro et al. (1996)]{NFW96} Navarro, J.F., Frenk, C.S. \& White, 
S.D.M. 1996, \apj, 462, 563
\bibitem[Nelan et al. (2005)]{Nel05} Nelan, J.E. et al. 2005, \apj, 632, 137
\bibitem[Neumann et al. (2001)]{Neu01} Neumann, D.M. et al. 2001, \aap, 365, L74
\bibitem[Pavlovsky et al. (2007)]{Pav07} Pavlovsky, C., et al. 2006, 
``Advanced Camera for Surveys Instrument Handbook for Cycle 16'', Version 7.1, 
(Baltimore: STScI).
\bibitem[Peng et al. (2006a)]{Peng06a} Peng, E.W. et al. 2006a, \apj, 639, 95
\bibitem[Peng et al. (2006b)]{Peng06b} Peng, E.W. et al. 2006a, \apj, 639, 838
\bibitem[Phillipps et al. (1998)]{Phi98} Phillipps, S., Driver, S.P., Couch, W.J
\& Smith, R.M. 1998, \apjl, 498, L119
\bibitem[Phillipps et al. (2001)]{Phi01} Phillipps, S., Drinkwater, M.J., Gregg, M. 
\& Jones, J.B. 2001, \apj, 560, 201
\bibitem[Pipino et al.(2007)]{Pip07} Pipino, A., Puzia, 
T.~H., \& Matteucci, F.\ 2007, ApJ in press, ArXiv e-prints, 704, arXiv:0704.0535
\bibitem[Pohlen \& Trujillo(2006)]{pohlen-trujillo} Pohlen, M., \& Trujillo,
I.\ 2006, \aap, 454, 759
\bibitem[Poggianti et al. (2001)]{Pog01} Poggianti, B.M. et al. 2001,
\apj, 562, 689
\bibitem[Poggianti et al. (2004)]{Pog04} Poggianti, B.M. et al. 2004,
\apj, 601, 197
\bibitem[Puzia et al. (2002)]{Puz02} Puzia, T.H., Zepf, S.E., Kissler-Patig, M., 
Hilker, M., Minniti, D. \& Goudfrooij, P. 2002, \aap, 391, 453
\bibitem[Rakos \& Schombert (2004)]{Rak04} Rakos, K. \& Schombert, J. 2004,
\aj, 127, 1502
\bibitem[Redmount \& Rees (1989)]{RR89} Redmount, I.~H., \& 
Rees, M.~J.\ 1989, Comments on Astrophysics, 14, 165  
\bibitem[Renaud et al. (2006)]{Ren06} Renaud, M., B\'elanger, G., Paul, J.,
Lebrun, F. \& Terrier, R. 2006, \aap 453, L5
\bibitem[Roberts et al. (2004)]{Rob04} Roberts, S. et al. 2004, \mnras, 352, 478
\bibitem[Roche et al. (1997)]{Roc97} Roche, N., Ratnatunga, K., Griffiths, R.E.
\& Im, M. 1997, \mnras, 288, 200
\bibitem[Sabatini et al. (2007)]{Sab07} Sabatini, S., Davies, J.I., Roberts. S.
\& Scaramella, R. 2007,  in ``Dark Galaxies and lost Baryons'', Proc.
I.A.U. symposium 244, eds: J.I. Davies \& M.J. Disney, (astro-ph/0707.4079)
\bibitem[S\'anchez-Bl\'azquez et al. (2006)]{Sanch06} S\'anchez-Bl\'azquez, P.,
Gorgas, J. \& Cardiel, N. 2006, \aap, 457, 823
\bibitem[Sembach et al. (2006)]{Sem06} Sembach, K. R., et al. 2006, 
``HST Two-Gyro Handbook'', Version 3.0, (Baltimore: STScI)
\bibitem[S\'ersic (1968)]{Sersic} S\`ersic, J.L. 1968, Atlas de Galaxias
Australes (C\'ordoba: Obs Astron., Univ. Nac. C\'ordoba)
\bibitem[Sheth et al.(2003)]{sheth03} Sheth, K., Regan, M. W., Scoville, N.
Z., \& Strubbe, L. E., 2003, \apj, 592, L13
\bibitem[Sirianni et al. (2003)]{Sir03} Sirianni, M., Martel, A. R., Jee,
  M. J., Van Orsow, D., \& Sparks, W. B.\ 2003, in ``2002 HST
  Calibration Workshop'', eds.\ S. Arribas, A. Koekemoer, \& B. Whitmore
  (Baltimore: STScI), 82
\bibitem[Sirianni et al. (2005)]{Sir05} Sirianni, M., et al. 2005, \pasp, 117, 1049.
\bibitem[Smith et al. (1997)]{Smi97} Smith, R.M., Driver, S.P., \& Phillipps, S.
1997, \mnras, 287, 415
\bibitem[Smith et al. (2004)]{Smi04} Smith, R.J. et al. 2004, \aj, 128, 1558
\bibitem[Smith et al. (2006)]{Smi06} Smith, R.J. Hudson, M.J., Lucey, J.R.,
Nelan J.E. \& Wegner, G.A. 2006, \mnras, 369, 1419
\bibitem[Somerville \& Primack (1999)]{Som99} Somerville, R.S. \& Primack, J.R.
1999, \mnras, 310, 1087
\bibitem[Thomsen et al. (1997)]{Tho97} Thomsen, B., Baum, W.A., Hammergren, M. \&
Worthey, G. 1997, \apjl, 483, L37
\bibitem[Tran et al. (2005)]{Tran05} Tran, K.-V.H., van Dokkum, P., Illingworth, G.D.,
Kelson, D., Gonzalez, A. \& Franx, M. 2005, \apj, 619, 134
\bibitem{Tre95} Tremaine, S.\ 1995, \aj, 110, 628 
\bibitem[Trentham et al. (2001)]{Tre01} Trentham, N., Tully, R.B., \& 
Verheijen, M.A.W. 2001, \mnras, 325, 385.
\bibitem[Trentham \& Hodgkin (2002)]{TH02} Trentham, N. \& Hodgkin, S.
2002, \mnras, 333, 423
\bibitem[Trentham \& Tully (2002)]{TT02} Trentham, N. \& Tully, R.B. 2002,
\mnras, 335, 712
\bibitem[Trujillo et al. (2004)]{Tru04} Trujillo, I., Erwin, P., Asensio Ramos,
A. \& Graham, A.W. 2004, \aj, 127, 1917
\bibitem[Tully \& Pierce (2000)]{Tul00} Tully, R.B. \& Pierce, M.J. 2000, \apj,
533, 744 
\bibitem[Tully et al. (2002)]{Tul02} Tully, R.B., Somerville, R., Trentham, N.
\& Verheijen, M.A.W. 2002, \apj, 569, 573
\bibitem[Valentijn et al. (2006)]{val06} Valentijn, E.A. et al. 2006, 
in ADASS XVI ASP Conference Series, 2006, Eds: R. Shaw, F.Hill and D. Bell
\bibitem[Valentijn \& Verdoes Kleijn (2006)]{VV06} Valentijn, E.A. \& Verdoes
Kleijn, G. 2006, ERCIM News, 65, 20
\bibitem[van den Bergh (2002)]{vdb02} van den Bergh, S. 2002, \aj, 124, 782
\bibitem[van Dokkum (2001)]{vandokkum01} van Dokkum, P. G.\ 2001,
  \pasp, 113, 1420
\bibitem[van Zee et al. (2004)]{vanZ04} van Zee, L., Barton, E. \& Skillman,
E.D. 2004, \aj, 128, 2797
\bibitem[Varela et al. (2004)]{Var04} Varela, J., Moles, M., M\'arquez, I., 
Galletta, G., Masegosa, J. \& Bettoni, D. 2004, \aa 420, 873
\bibitem[Vogt et al. (2004)]{Vog04} Vogt, N.P., Haynes, M.P., Herter, T.,
\& Giovanelli, R. 2004, \aj, 127, 3273
\bibitem[White et al. (1993)]{Whi93} White, S.D.M., Briel, U.G. \& Henry, J.P
1993, \mnras, 261, L8
\bibitem[Wirth \& Gallagher (1984)]{Wir84} Wirth, A. \& Gallagher, J.S. 1984,
\apj, 282, 85
\bibitem[Woodworth \& Harris (2000)]{Woo00} Woodworth, S.C. \& Harris, W.E.
2000, \aj, 119, 2699
\bibitem[Younger et al.(2007)]{younger07} Younger, J. D., Cox, T. J., Seth, A.
C., \& Hernquist, L. 2007, \apj, in press
\bibitem[Zheng et al. (2005)]{Zhe05} Zheng, X.Z., Hammer, F., Flores, H., 
Ass\'emat, F. \& Rawat, A. 2005, \aap, 435, 507
\bibitem[Zwaan et al. (2005)]{Zwa05} Zwaan, M.A., Meyer, M.J., Stavely-Smith, L.
\& Webster, R.L. 2005, \mnras, 359, 30

\end{thebibliography}
\end{document}